\documentclass[journal]{IEEEtran}            %IEEE Transactions format. Proof env needs to be redefined. 

\IEEEoverridecommandlockouts                              % This command is only needed if 
\usepackage[utf8]{inputenc}
\usepackage[english]{babel}
\usepackage[normalem]{ulem}
\usepackage{amsmath}
\usepackage{amssymb}
\usepackage{amsfonts}
\usepackage{centernot}
\usepackage{fouriernc}
\usepackage{algorithm}
\usepackage{algorithmic}
\usepackage{graphicx}
\usepackage{hyperref}
\newtheorem{thm}{Theorem}[section]
\newtheorem{lm}[thm]{Lemma}
\newtheorem{cor}[thm]{Corollary}
\newtheorem{df}[thm]{Definition}
\newtheorem{prob}[thm]{Problem}

\newtheorem{prop}[thm]{Proposition}
\newtheorem{eg}[thm]{Example}

\usepackage[utf8]{inputenc}
\usepackage[english]{babel}
 \usepackage{graphicx}
\usepackage{subcaption}
\usepackage[usenames, dvipsnames]{color}
\definecolor{red}{rgb}{1,0.2,0.2}

\renewcommand{\algorithmiccomment}[1]{\hskip4em$\triangleright$ #1}
\usepackage{amssymb}
\title{Secure Control in Partially Observable Environments to Satisfy LTL Specifications$^{*,+}$}
\author{Bhaskar Ramasubramanian$^{1}$, Luyao Niu$^{2}$, Andrew Clark$^{2}$,\\ Linda Bushnell$^{1}$, \emph{Fellow, IEEE}, and Radha Poovendran$^{1}$,  \emph{Fellow, IEEE}% <-this % stops a space
\thanks{$^*$This work was supported by the U.S. Army Research Office, the National Science Foundation, and the Office of Naval Research via Grants W911NF-16-1-0485, CNS-1941670, and N00014-17-S-B001 respectively.}
\thanks{$^+$A preliminary version of this work appears in \cite{bhaskar2019finite}.}
\thanks{$^{1}$Network Security Lab, Department of Electrical and Computer Engineering, 
University of Washington, Seattle, WA 98195, USA. \newline
        {\tt\small \{bhaskarr, lb2, rp3\}@uw.edu}}%
\thanks{$^{2}$Department of Electrical and Computer Engineering, 
Worcester Polytechnic Institute, Worcester, MA 01609, USA.  \newline
        {\tt\small \{lniu, aclark\}@wpi.edu}}%
}
\date{}
\begin{document}
\maketitle
\begin{abstract} 

This paper studies the synthesis of control policies for an agent that has to satisfy a temporal logic specification in a partially observable environment, in the presence of an adversary. 
The interaction of the agent (defender) with the adversary is modeled as a partially observable stochastic game. 
The goal is to generate a defender policy to maximize satisfaction of a given temporal logic specification under any adversary policy. 
The search for policies is limited to the space of finite state controllers, which leads to a tractable approach to determine policies. 
We relate the satisfaction of the specification to reaching (a subset of) recurrent states of a Markov chain. 
We present an algorithm to determine a set of defender and adversary finite state controllers of fixed sizes that will satisfy the temporal logic specification, and prove that it is sound. 
We then propose a value-iteration algorithm to maximize the probability of satisfying the temporal logic specification under finite state controllers of fixed sizes. 
Lastly, we extend this setting to the scenario where the size of the finite state controller of the defender can be increased to improve the satisfaction probability. 
We illustrate our approach with an example. 
\end{abstract}
%%%%%%%%%%%%%%%%%%%%%%%%%%%%%%%%%%%%%%%%%%%%%%%%%%%%%%%%%
%
\section{Introduction}\label{Introduction}

Cyber-physical systems (CPSs) are entities in which the working of a physical system is governed by its interactions with computing devices and algorithms. 
These systems are ubiquitous \cite{baheti2011cyber}, and vary in scale from energy systems to medical devices and robots. 
In applications like autonomous cars and robotics, CPSs are expected to operate in dynamic and potentially dangerous environments with a large degree of autonomy. 
In such a setting, the system might be the target of malicious attacks that aim to prevent it from accomplishing a goal. 
An attack can be carried out on the physical system, on the computers that control the physical system, or on communication channels between components of the system. 
Such attacks by an intelligent attacker have been reported across multiple domains, including power systems \cite{sullivan2017cyber}, automobiles \cite{shoukry2013non}, water networks \cite{slay2007lessons}, and nuclear reactors \cite{farwell2011stuxnet}. 
%It is often the case that adversaries are stealthy and tailor their attacks to cause maximum damage. 
Adversaries are often stealthy, and tailor their attacks to cause maximum damage. 
Therefore, strategies designed to only address modeling and sensing errors may not satisfy performance requirements in the presence of an intelligent adversary who can manipulate system operation. 
%A natural question to ask before solving a problem in this domain is the means by which the environment, goals, and constraints, if any, are specified. 

The preceding discussion makes it imperative to develop methods to specify and verify CPSs and the environments they operate in. 
Formal methods \cite{baier2008principles} enable the verification of the behavior of CPS models against a rich set of specifications \cite{lahijanian2015formal}. 
Properties like safety, liveness, stability, and priority can be expressed as formulas in linear temporal logic (LTL) \cite{kress2007s, ding2014optimal}, and can be 
%Linear temporal logic (LTL) is a particularly well-understood framework to express properties like safety, liveness, and priority \cite{kress2007s, ding2014optimal}. 
%These properties can be 
verified using off-the-shelf model solvers \cite{cimatti1999nusmv, kwiatkowska2011prism} that take these formulas as inputs. 
Markov decision processes (MDPs) \cite{bertsekas2015dynamic, puterman2014markov} have been used to model environments where outcomes depend on both, an inherent randomness in the model (transition probabilities) and an action taken by an agent.  
These models have been extensively used in applications like robotics \cite{lahijanian2012temporal} and unmanned aircrafts \cite{temizer2010collision}. 

Current literature on the satisfaction of an LTL formula over an MDP assumes that states are fully observable \cite{ding2014optimal, lahijanian2012temporal, niu2018secure}. 
In many practical scenarios, states may not be observable. 
For example, as seen in \cite{thrun2005probabilistic}, a robot might only have an estimate of its location based on the output of a vision sensor. 
The inability to observe all states necessitates the use of a framework that accounts for partial observability. 
For LTL formula satisfaction in partially observable environments with a single agent, partially-observable Markov decision processes (POMDPs) can be used to model and solve the problem \cite{sharan2014finite, sharan2014formal}. 
However, determining an `optimal policy' for an agent in a partially observable environment is NP-hard for the infinite horizon case, which was shown in \cite{vlassis2012computational}. 
This demonstrates the need for techniques to determine approximate solutions. 

Heuristics to approximately solve POMDPs include belief replanning \cite{cassandra1996acting}, most likely belief state policy and entropy weighting \cite{kaelbling1998planning}, grid-based methods \cite{brafman1997heuristic}, and point-based methods \cite{kurniawati2008sarsop}. 
The difficulty in computing exactly optimal policies and the lack of complete observability may be exploited by an adversary to launch new attacks on the system. 
The synthesis of parameterized finite state controllers (FSCs) for a POMDP to maximize the probability of satisfying of an LTL formula (in the absence of an adversary) was proposed in \cite{sharan2014finite} and \cite{sharan2014formal}. 
This is an approximate strategy since it does not use the observation and action histories; it uses only the most recent observation in order to determine an action. 
This restricts the class of policies that are searched over, 
but the finite cardinality of states in an FSC makes the problem computationally tractable. 
The authors of \cite{yu2008near} showed the existence of $\epsilon-$optimal FSCs for the average cost POMDP. 
%However, 
%In lieu of 
%a detailed study of the quantification of this approximation in our setting where we have two agents,  
%is beyond the scope of this paper. 
%Instead
In comparison, for the setting in this paper where we have two competing agents, 
we present guarantees on the convergence of a value-iteration based procedure in terms of the number of states in the environment and the FSCs.
%Moreover, an adversary might be able to exploit the lack of complete observability of the environment and the ability to determine optimal policies to launch more damaging attacks on the system. 

In this paper, we study the problem of determining strategies for an agent that has to satisfy an LTL formula in the presence of an adversary in a partially observable environment. 
The agent and the adversary take actions simultaneously, and these jointly influence transitions between states. % of the system. 
%
%, and in particular, can be used to solve the average reward problem over the infinite horizon \cite{sharan2014formal}. 
%
\subsection{Contributions}

The setting that we consider in this paper assumes two players or agents-- a defender and an adversary-- who are each limited in that they do not exactly observe the state. 
%We extend the setting of \cite{sharan2014formal} to include an adversary who is also limited in that it does not exactly observe the state. 
The policies of the agents are represented as FSCs. %, whose goals are opposite to each other. 
The goal for the defender will be to synthesize a policy that will maximize the probability of satisfying an LTL formula for any adversary policy. 
%The adversary policy is also represented by an FSC, whose goal is opposite to that of the defender. 
We make the following contributions. 
\begin{itemize}
\item We show that maximizing the satisfaction probability of the LTL formula under any adversary policy is equivalent to maximizing the probability of reaching a recurrent set of a Markov chain constructed by composing representations of the environment, the LTL objective, and the respective agents' controllers.% This Markov chain additionally contains states that need to be reached in order to satisfy the LTL formula. 

\item %The search for policies involves optimizing over both the sizes of the FSCs and their transition probabilities. 
We develop a heuristic algorithm to determine defender and adversary FSCs of fixed sizes that will satisfy the LTL formula with nonzero probability, and show that it is sound. 
%We present a \textcolor{blue}{sound algorithm to} determine defender and adversary FSCs of fixed sizes that will satisfy the LTL formula with nonzero probability. 
The search for a defender policy that will maximize the probability of satisfaction of the LTL formula for any adversary policy can then be reduced to a search among these FSCs.% of fixed size. 
%If these FSCs are parameterized in an appropriate way, it might lend itself to gradient-based optimization techniques. 

\item We propose a procedure based on value-iteration 
%value-iteration \textcolor{blue}{based} procedure 
that maximizes the probability of satisfying the LTL formula under fixed defender and adversary FSCs. 
This satisfaction probability is related to a Stackelberg equilibrium of a partially observable stochastic game involving the defender and adversary. 
We also give guarantees on the convergence of this procedure.

\item We study the case when the size of the defender FSC can be changed to improve the satisfaction probability.

\item We present an example to illustrate our approach.
\end{itemize}

The value-iteration procedure and the varying defender FSC size described above is new to this work, along with more detailed examples. 
%We additionally present the other results in greater detail.  
This differentiates the present paper from a preliminary version that appears in \cite{bhaskar2019finite}. 
\subsection{Outline}

An overview of LTL and partially observable stochastic games (POSGs) is given in Section \ref{Prelim}. 
We define FSCs for the two agents, and show how they can be composed with a POSG to yield a Markov chain in Section \ref{Problem}. 
%Sections \ref{LTLRecSet}--\ref{FSCVary} present the main results of this paper. 
Section \ref{LTLRecSet} relates LTL satisfaction on a POSG to reaching specific subsets of recurrent sets of an associated Markov chain. 
Section \ref{FindFSCs} gives a procedure to determine %candidate 
defender and adversary FSCs of fixed sizes that will ensure that the LTL formula will be satisfied with non-zero probability. 
%This section also proposes a rule to determine the subset of states to visit in steady-state. 
A value-iteration procedure to maximize the probability of satisfying the LTL formula under fixed defender and adversary FSCs is detailed in Section \ref{POSGValIter}. 
Section \ref{FSCVary} addresses the scenario when states may be added to the defender FSC in order to improve the probability of satisfying the LTL formula under an adversary FSC of fixed size.  
Illustrative examples are presented in Section \ref{Example}. 
Section \ref{RelWork} summarizes related work in POMDPs and TL satisfaction on MDPs.  
Section \ref{Conclusion} concludes the paper. 
%Section \ref{Conclusion} presents concluding remarks.
%
\section{Preliminaries}\label{Prelim}

In this section, we give a concise introduction to linear temporal logic and partially observable stochastic games. 
We then detail the construction of an entity which will ensure that runs on a POSG will satisfy an LTL formula. 
\subsection{Linear Temporal Logic}\label{LTL}

Temporal logic frameworks enable the representation and reasoning about temporal information on propositional statements. 
\emph{Linear temporal logic (LTL)} is one such framework, where the progress of time is `linear'. 
An \emph{LTL formula} \cite{baier2008principles} is defined over a set of atomic propositions $\mathcal{AP}$, and can be written as: 
%\begin{align*}
$\phi:=\mathtt{T}|\sigma| \neg \phi | \phi \wedge \phi | \mathbf{X} \phi |\phi \mathbf{U} \phi$, 
%\end{align*}
%
where $\sigma \in \mathcal{AP}$, and $\mathbf{X}$ and $\mathbf{U}$ are temporal operators denoting the \emph{next} and \emph{until} operations respectively.  

The semantics of LTL are defined over (infinite) words in $2^{\mathcal{AP}}$. 
We write $\eta_0 \eta_1\dots:=\eta \models \phi$ when a trace $\eta \in (2^{\mathcal{AP}})^{\omega}$ satisfies an LTL formula $\phi$. 
Here, the superscript $\omega$ serves to indicate the potential infinite length of the word\footnote{To be more precise, $\eta$ is a word in an \emph{$\omega-$regular language}, which is a generalization of regular languages to words of infinite length \cite{baier2008principles}.}% A comprehensive treatment of this topic is beyond the scope of this paper, and we direct the interested reader to Chapter $4$ of \cite{baier2008principles}.}. 
\begin{df}[LTL Semantics]\label{LTLSemantics}
Let $\eta^i = \eta_i\eta_{i+1}\dots$. 
Then, the semantics of LTL can be recursively defined as: 
\begin{enumerate}
\item $\eta \models \mathtt{T}$ if and only if (iff) $\eta_0$ is true; 
\item $\eta \models \sigma$ iff $\sigma \in \eta_0$; 
\item $\eta \models \neg \phi$ iff $\eta \not \models \phi$; 
\item $\eta \models \phi_1 \wedge \phi_2$ iff $\eta \models \phi_1$ and $\eta \models \phi_2$; 
\item $\eta \models \mathbf{X} \phi$ iff $\eta^1 \models \phi$; 
\item $\eta \models \phi_1 \mathbf{U} \phi_2$ iff $\exists j \geq 0$ such that $\eta^j \models \phi_2$ and for all $k < j, \eta^k \models \phi_1$. 
\end{enumerate}
\end{df}

Moreover, the logic admits derived formulas of the form: 
\emph{i)} $\phi_1 \vee \phi_2:=\neg(\neg \phi_1 \wedge \neg \phi_2)$; 
\emph{ii)} $\phi_1 \Rightarrow \phi_2:= \neg \phi_1 \vee \phi_2$; 
\emph{iii)} $\mathbf{F}\phi:=\mathtt{T} \mathbf{U} \phi  \text{  (eventually)}$; 
\emph{iv)} $\mathbf{G} \phi:= \neg \mathbf{F} \neg \phi \text{  (always)}$. 
\begin{df}[Deterministic Rabin Automaton]
A \emph{deterministic Rabin automaton (DRA)} is a quintuple $\mathcal{RA} = (Q, \Sigma, \delta, q_0,F)$ where $Q$ is a nonempty finite set of states, $\Sigma$ is a finite alphabet, $\delta : Q \times \Sigma \rightarrow Q$ is a transition function, $q_0 \in Q$ is the initial state, and $F:=\{(L(i),K(i)\}_{i=1}^M$ is such that $L(i), K(i) \subseteq Q$ for all $i$, and $M$ is a positive integer. 
\end{df}

A \emph{run} of $\mathcal{RA}$ is an infinite sequence of states $q_0q_1\dots$ such that $q_{i} \in \delta (q_{i-1}, \alpha)$ for all $i$ and for some $\alpha \in \Sigma$. 
The run is \emph{accepting} if there exists $(L,K) \in F$ such that the run intersects with $L$ finitely many times, and with $K$ infinitely often. 
An LTL formula $\phi$ over $\mathcal{AP}$ can be represented by a DRA with alphabet $2^{\mathcal{AP}}$  that accepts all and only those runs that satisfy $\phi$. 
\subsection{Stochastic Games and Markov Chains}\label{SGMC}

A stochastic game involves one or more players, and starts with the system in a particular state. 
Transitions to a subsequent state are probabilistically determined by the current state and the actions chosen by each player, and this process is repeated. 
%In the sequel, we formally define a two-player stochastic game. 
%Since our focus will be on games of this type, 
Our focus will be on two-player stochastic games, and we omit the quantification on the number of players for the remainder of this paper. 
\begin{df}[Stochastic Game]\label{SGDefn}
A \emph{stochastic game} \cite{niu2018secure} is a tuple $\mathcal{G}:=(S, s_0, U_{d}, U_{a}, \mathbb{T}, \mathcal{AP}, \mathcal{L})$. 
$S$ is a finite set of states, $s_0$ is the initial state, $U_{d}$ and $U_{a}$ are finite sets of actions of the defender and adversary, respectively. 
%The function 
$\mathbb{T}: S \times U_{d} \times U_{a}  \times S \rightarrow [0,1]$ encodes $\mathbb{T}(s'|s,u_{d},u_{a})$, the transition probability from state $s$ to $s'$ when defender and adversary actions are $u_{d}$ and $u_{a}$, such that $\sum_{s'}\mathbb{T}(s'|s,u_{d},u_{a}) =1$ for all $s,u_{d},u_{a}$. 
$\mathcal{AP}$ is a set of atomic propositions. $\mathcal{L}: S \rightarrow 2^{\mathcal{AP}}$ is a labeling function that maps a state to a subset of atomic propositions that are satisfied in that state. 
\end{df}

Stochastic games can be viewed as an extension of Markov Decision Processes when there is more than one player taking an action. 
For a player, a \emph{policy} is a mapping from sequences of states to actions, if it is deterministic, or from sequences of states to a probability distribution over actions, if it is randomized. 
A policy is \emph{Markov}
%\emph{stationary} 
if it is dependent only on the most recent state. 

In this paper, we focus our attention on the Stackelberg setting \cite{fudenberg1991game}. 
In this framework, the first player (leader) commits to a policy. 
The second player (follower) observes the leader's policy and chooses its policy as the best response to the leader's policy, defined as the policy that maximizes the follower's utility. 
We also assume that the players take their actions concurrently at each time step. 
%The reader is referred to \cite{niu2018optimal} for an example that shows the advantage of the concurrent setting with agents playing randomized policies, over a turn-based setting and/ or deterministic policies. 

We now define the notion of a Stackelberg equilibrium, which indicates that a solution to a Stackelberg game has been found. 
%Let the leader's (follower's) policy be $\mathit{l}$ ($\mathit{f}$), and let  
%Denote the leader's policy by $\mathit{l}$ and the follower's policy by $\mathit{f}$. 
Let $Q_L(\mathit{l}, \mathit{f})$ ($Q_F(\mathit{l}, \mathit{f})$) be the utility gained by the leader (follower) by adopting a policy $\mathit{l}$ ($\mathit{f}$). 
\begin{df}[Stackelberg Equilibrium]\label{StackEq}
A pair $(\mathit{l}, \mathit{f})$ is a \emph{Stackelberg equilibrium} if $\mathit{l} = \arg \max_{\mathit{l}'}Q_L(\mathit{l}', BR(\mathit{l}'))$, where $BR(\mathit{l}') = \{\mathit{f}: \mathit{f}=\arg \max Q_F(\mathit{l}',\mathit{f})\}$. 
That is, the leader's policy is optimal given that the follower observes the leader's policy and plays its best response. 
\end{df}

%We end this subsection with a primer on terms from Markov Chains that will be used in this paper. 
%When $U_{a} = \emptyset$ and $|U_{d}| = 1$, 
When $|U_a| = |U_d| = 1$, 
$\mathcal{G}$ is a \emph{Markov chain} \cite{meyn2012markov}. 
For $s, s' \in S$, $s'$ is \emph{accessible} from $s$, written $s \rightarrow s'$, if %the transition probability from $s$ to $s'$, given by 
$\mathbb{P}(s_a|s) \mathbb{P}(s_b|s_a) \dots \mathbb{P}(s_i|s_j) \mathbb{P}(s'|s_i) > 0$ for some (finite subset of) states $s_a,s_b,\dots,s_i,s_j$. 
%Equivalently, $s \rightarrow s'$ if there is a positive probability of reaching $s'$ from $s$ in a finite number of steps. 
Two states \emph{communicate} %, written $s \leftrightarrow s'$, 
if $s \rightarrow s'$ and $s' \rightarrow s$. 
\emph{Communicating classes} of states cover the state space of the Markov chain. 
A state is \emph{transient} if there is a nonzero probability of not returning to it when we start from that state, and is \emph{positive recurrent} otherwise. 
%If some state in a communicating class is recurrent (transient), then the same holds for all other states in that class. 
In a finite state Markov chain, every state is either transient or positive recurrent.
%We refer the reader to \cite{meyn2012markov} for a detailed exposition. 
%
\subsection{Partially Observable Stochastic Games}\label{POSG}

%The states are fully observable in Definition \ref{SGDefn}. 
Partially observable stochastic games (POSGs) extend Definition \ref{SGDefn} to the case when instead of observing a state directly, each player receives an observation that is derived from the state.  
This can be viewed as an extension of POMDPs to the case when there is more than one player. 
\begin{df}[Partially Observable Stochastic Game]
A \emph{partially observable stochastic game} is defined by the tuple $\mathcal{SG} := (S, s_0, U_{d}, U_{a}, \mathbb{T}, \mathcal{O}_{d}, \mathcal{O}_{a}, O_{d}, O_{a}, \mathcal{AP}, \mathcal{L})$, where $S, s_0, U_{d}, U_{a}, \mathbb{T}, \mathcal{AP}, \mathcal{L}$ are as in Definition \ref{SGDefn}. $\mathcal{O}_{d}$ and $\mathcal{O}_{a}$ denote the (finite) sets of observations available to the defender and adversary. $O_*:S \times \mathcal{O}_* \rightarrow [0,1]$ encodes $\mathbb{P}(o_*|s)$, where $* \in \{d, a\}$. 
\end{df}

The functions $O_*$ model imperfect sensing. 
In order for $O_*$ to satisfy the conditions of a probability distribution, we need $O_*(o|s) \geq 0 \forall o \in \mathcal{O}_*$ and $\sum_{o \in \mathcal{O}_*} O_*(o|s) = 1$. 

\subsection{Adversary and Defender Models}

The initial state of the system is $s_0$. 
A transition from a state $s_t$ to the next state $s_{t+1}$ is determined jointly by the actions of the defender and adversary according to the transition probability function $\mathbb{T}$.

At a state $s_t$, the adversary makes an observation, $O_a^t$ of the state according to $O_a$. 
The adversary is also assumed to be aware of the policy (sequence of actions), $\mu_d$, committed to by the defender. 
Therefore, the overall information available to the adversary is $\mathfrak{I}^t_a := \bigcup \limits_{i=0:t} O_a^i \cup \{\mu_d\}$.

Different from the information available to the adversary, at state $s_t$, the defender makes an observation $O_d^t$ of the state according to $O_d$. 
Therefore, the overall information for the defender is $\mathfrak{I}^t_d := \bigcup \limits_{i=0:t} O_d^i$.

%\textcolor{blue}{At each time step $t$, the adversary receives an observation, $O_a^t$ of the state according to $O_a$. 
%The adversary is also aware of the policy (sequence of inputs), $\mu_d$, committed to by the defender. 
%Therefore, the overall information available to the adversary is $\mathfrak{I}^t_a := \bigcup \limits_{i=0:t} O_a^i \cup \{\mu_d\}$. 
%On the other hand, at time $t$, the information available to the defender is only in terms of the observations of the state, $O_d^t$, which it receives according to $O_d$. 
%Therefore, the overall information for the defender is $\mathfrak{I}^t_d := \bigcup \limits_{i=0:t} O_d^i$.}
%
%The information available until time $t$, denoted $\mathfrak{I}_t$, can be inductively defined as: $\mathfrak{I}_0 = S$, $\mathfrak{I}_t = \mathfrak{I}_{t-1} \times U_{d} \times \mathcal{O}_{d} \times U_{a} \times \mathcal{O}_{a}$. 
%The overall information is $\mathfrak{I} := \cup_t \mathfrak{I}_t$. 
%
\begin{df}[POSG Policy]
A \emph{(defender or adversary) policy} for the POSG is a map from the respective overall information to a probability distribution over the corresponding action space, i.e. $\mu^t_*: \mathfrak{I}^t_* \times U_* \rightarrow [0,1]$, $* \in \{d, a\}$. 
\end{df}

Policies of the form above are called \emph{randomized policies}. 
If $\mu^t_*:\mathfrak{I}^t_*\rightarrow U_*$, it is called a \emph{deterministic policy}. 
In the sequel, we will use finite state controllers as a representation of policies that consider only the most recent observation.
\subsection{The Product-POSG}\label{ProdPOSG}

In order to find runs on $\mathcal{SG}$ that would be accepted by a DRA $\mathcal{RA}$ built from an LTL formula $\phi$, we construct a product-POSG. 
This construction is motivated by the product-stochastic game construction in \cite{niu2018secure} and the product-POMDP construction in \cite{sharan2014finite}. 

\begin{figure*}
% ensure that we have normalsize text
\normalsize
% Store the current equation number.
%\setcounter{mytempeqncnt}{\value{equation}}
% Set the equation number to one less than the one
% desired for the first equation here.
% The value here will have to changed if equations
% are added or removed prior to the place these
% equations are referenced in the main text.
%\setcounter{equation}{5}
\begin{align}
\bar{\mathbb{T}}:=&\mathbb{T}^{\phi, \mathcal{C}_{d}, \mathcal{C}_{a}}((s',q'),g'_{d},g'_{a}|(s,q),g_{d},g_{a}) \label{TransFn} \\ &= \sum_{o \in \mathcal{O}_{d}} \sum_{o' \in \mathcal{O}_{a}} \sum_{u_{d}} \sum_{u_{a}} O_{d}(o|s) O_{a}(o'|s) \mu_{d}(g'_{d}, u_{d}|g_{d}, o) \mu_{a}(g'_{a}, u_{a}|g_{a}, o') \mathbb{T}^{\phi}((s',q')|(s,q),u_{d}, u_{a})\nonumber
\end{align}
% Restore the current equation number.
%\setcounter{equation}{\value{mytempeqncnt}}
% IEEE uses as a separator
\hrulefill
% The spacer can be tweaked to stop underfull vboxes.
%\vspace*{2pt}
\end{figure*}
\begin{df}[Product-POSG]
Given $\mathcal{SG}$ and $\mathcal{RA}$ (built from LTL formula $\phi$), a \emph{product-POSG} is $\mathcal{SG}^{\phi} = (S^{\phi}, s_0^\phi,U_{d}, U_{a}, \mathbb{T}^{\phi}, \mathcal{O}_{d}, \mathcal{O}_{a}, O_{d}^{\phi}, O_{a}^{\phi}, F^{\phi}, \mathcal{AP}, \mathcal{L}^{\phi})$, 
where $S^{\phi} = S \times Q$, $s_0^\phi = (s_0, q_0)$, $O_{*}^{\phi}(o|(s,q)) = O_{*}(o|s)$, $\mathbb{T}^{\phi}((s',q')|(s,q),u_{d}, u_{a}) = \mathbb{T}(s'|s,u_{d}, u_{a})$ iff $\delta(q,\mathcal{L}(s')) = q'$, and $0$ otherwise, 
%$O_{d}^{\phi}(o|(s,q)) = O_{d}(o|s)$, $O_{a}^{\phi}(o|(s,q)) = O_{a}(o|s)$,
$F^{\phi} = \{(L^{\phi}(i), K^{\phi}(i))\}_{i=1}^M$, $L^{\phi}(i), K^{\phi}(i) \subset S^{\phi}$, and $(s,q) \in L^{\phi}(i)$ iff $q \in L(i)$, $(s,q) \in K^{\phi}(i)$ iff $q \in K(i)$, $\mathcal{L}^{\phi}((s,q)) = \mathcal{L}(s)$.
\end{df}

From the above definition, it is clear that acceptance conditions in the product-POSG depend on the DRA while the transition probabilities of the product-POSG are determined by transition probabilities of the original POSG. 
Therefore, a run on the product-POSG can be used to generate a path on the POSG and a run on the DRA. 
Then, if the run on the DRA is accepting, we say that the product-POSG satisfies the LTL specification $\phi$. 
\section{Problem Setup}\label{Problem}

This section details the construction of FSCs for the two agents. 
An FSC for an agent can be interpreted as a policy for that agent. 
The defender and adversary policies will be determined by probability distributions over transitions in finite state controllers that are composed with the POSG. 
%This method is chosen because the FSCs when composed with the product-POSG, will result in a finite state Markov chain. 
When the FSCs are composed with the product-POSG, the resulting entity is a Markov chain. 
We then establish a way to determine satisfaction of an LTL specification on the product-POSG in terms of runs on the composed MC. 
%A treatment for the single-agent case when the environment is specified as a POMDP was presented in \cite{sharan2014finite}. 
%
%\begin{figure*}
%% ensure that we have normalsize text
%\normalsize
%% Store the current equation number.
%%\setcounter{mytempeqncnt}{\value{equation}}
%% Set the equation number to one less than the one
%% desired for the first equation here.
%% The value here will have to changed if equations
%% are added or removed prior to the place these
%% equations are referenced in the main text.
%%\setcounter{equation}{5}
%\begin{align}
%\bar{\mathbb{T}}:=&\mathbb{T}^{\phi, \mathcal{C}_{d}, \mathcal{C}_{a}}((s',q'),g'_{d},g'_{a}|(s,q),g_{d},g_{a}) \label{TransFn} \\ &= \sum_{o \in \mathcal{O}_{d}} \sum_{o' \in \mathcal{O}_{a}} \sum_{u_{d}} \sum_{u_{a}} O_{d}(o|s) O_{a}(o'|s) \mu_{d}(g'_{d}, u_{d}|g_{d}, o) \mu_{a}(g'_{a}, u_{a}|g_{a}, o') \mathbb{T}^{\phi}((s',q')|(s,q),u_{d}, u_{a})\nonumber
%\end{align}
%% Restore the current equation number.
%%\setcounter{equation}{\value{mytempeqncnt}}
%% IEEE uses as a separator
%\hrulefill
%% The spacer can be tweaked to stop underfull vboxes.
%%\vspace*{2pt}
%\end{figure*}
%
\subsection{Finite State Controllers}\label{FSCs}

Finite state controllers consist of a finite number of internal states. 
Transitions between states is governed by the current observation of the agent. 
%A directed cyclic graph of internal states of the FSC will allow for remembering events relevant to taking optimal actions \cite{sharan2014finite}. 
In our setting, we will have two FSCs, one for the defender and another for the adversary. 
We will then limit the search for defender and adversary policies to one over FSCs of fixed cardinality. 
\begin{df}[Finite State Controller]
A \emph{finite state controller for the defender (adversary)}, denoted $\mathcal{C}_{d}$ ($\mathcal{C}_{a}$), is a tuple $\mathcal{C}_* = (G_*,g_{0_*}, \mu_*)$, where $G_*$ is a finite set of (internal) states of the controller, $g_{0_*}$ is the initial state of the FSC, and $\mu_*: G_* \times \mathcal{O}_* \times G_* \times U_* \rightarrow [0,1]$, written $\mu_*(g'_*, u_*|g_*, o_*)$, is a probability distribution of the next internal state and action, given a current internal state and observation. 
%The initial state of $\mathcal{C}_*$ is a probability distribution over $G_*$, and will depend on the initial state of the system. 
Here, $* \in \{d, a\}$. 
\end{df}

An FSC is a finite-state probabilistic automaton that takes the current observation of the agent as its input, and produces a distribution over the actions as its output. 
The FSC-based control policy is defined as follows: initial states of the FSCs are determined by the initial state of the POSG. 
The defender commits to a policy at the start. 
At each time step, the policy returns a distribution over the actions and the next state of $\mathcal{C}_{d}$, given the current state of the FSC $\mathcal{C}_{d}$ and the state of $\mathcal{SG}^{\phi}$ observed according to $O_{d}$. 
The adversary observes this and the state according to $O_{a}$ and responds with $\mu_{a}(\cdot)$ generated by $\mathcal{C}_{a}$. 
Actions at each step are taken concurrently.  
\begin{df}[Proper FSCs]\label{FSCProper}
An FSC is \emph{proper} with respect to an LTL formula $\phi$ if there is a positive probability of satisfying $\phi$ under this policy in an environment represented as a partially observable stochastic game. 
\end{df}

This is similar to the definition in \cite{hansen2003synthesis}, with the distinction that the terminal state of an FSC in that context will be related to Rabin acceptance pairs of an MC formed by composing $\mathcal{C}_{d}$ and $\mathcal{C}_{a}$ with $\mathcal{SG}^{\phi}$ (Sec \ref{GMC}). 
%We will restrict ourselves to proper FSCs for the rest of this paper <MOVE TO END OF SEC V>. 
%
\subsection{The Global Markov Chain}\label{GMC}

The FSCs $\mathcal{C}_{d}$ and $\mathcal{C}_{a}$, when composed with $\mathcal{SG}^{\phi}$, will result in a finite-state, (fully observable) Markov chain. 
To maintain consistency with the literature, we will refer to this as the \emph{global Markov chain (GMC)} \cite{sharan2014finite}. 
%In the following definition, we use the notation $\bar{W} := W^{\phi, \mathcal{C}_{d}, \mathcal{C}_{a}}$ to denote the entity $W$ resulting from a product-POSG $\mathcal{SG}^{\phi}$ controlled by FSCs $\mathcal{C}_{d}$ and $\mathcal{C}_{a}$.
%
%\begin{df}[Global Markov Chain]
%The \emph{global Markov chain} resulting from a product-POSG $\mathcal{SG}^{\phi}$ controlled by FSCs $\mathcal{C}_{d}$ and $\mathcal{C}_{a}$ is $\mathcal{M}:=\mathcal{M}^{\phi, \mathcal{C}_{d}, \mathcal{C}_{a}} = (S^{\phi, \mathcal{C}_{d}, \mathcal{C}_{a}}, s_0^{\phi, \mathcal{C}_{d}, \mathcal{C}_{a}},\mathbb{T}^{\phi, \mathcal{C}_{d}, \mathcal{C}_{a}}, \mathcal{AP}, \mathcal{L}^{\phi, \mathcal{C}_{d}, \mathcal{C}_{a}})$. $S^{\phi, \mathcal{C}_{d}, \mathcal{C}_{a}} = $ $S^{\phi} \times G_{d} \times G_{a}$, 
%$s_0^{\phi, \mathcal{C}_{d}, \mathcal{C}_{a}} = (s_0, q_0, g_{0_{d}}, g_{0_{a}})$, 
%$\mathbb{T}^{\phi, \mathcal{C}_{d}, \mathcal{C}_{a}}$ is given by Equation (\ref{TransFn}), and $\mathcal{L}^{\phi, \mathcal{C}_{d}, \mathcal{C}_{a}}((s,q),g_{d}, g_{a}) =\mathcal{L}^{\phi}((s,q))$.
%\end{df}
\begin{df}[Global Markov Chain (GMC)]
The \emph{GMC} resulting from a product-POSG $\mathcal{SG}^{\phi}$ controlled by FSCs $\mathcal{C}_{d}$ and $\mathcal{C}_{a}$ is $\mathcal{M}:=\mathcal{M}^{\phi, \mathcal{C}_{d}, \mathcal{C}_{a}} = (\bar{S}, \bar{s}_0,\bar{\mathbb{T}}, \mathcal{AP}, \bar{\mathcal{L}})$, where $\bar{S}= $ $S^{\phi} \times G_{d} \times G_{a}$, 
$\bar{s}_0= (s_0, q_0, g_{0_{d}}, g_{0_{a}})$, 
$\bar{\mathbb{T}}$ is given by Equation (\ref{TransFn}), and $ \bar{\mathcal{L}} =\mathcal{L}^{\phi}((s,q))$.
\end{df}

Similar to $\mathcal{SG}^{\phi}$, the Rabin acceptance condition for $\bar{\mathcal{M}}$ is: $\bar{F}= \{(\bar{L}(i), \bar{K}(i))\}_{i=1}^M$, with $(s,q,g_{d},g_{a}) \in \bar{L}(i)$ iff $(s,q) \in L^{\phi}(i)$ and $(s,q,g_{d},g_{a}) \in \bar{K}(i)$ iff $(s,q) \in K^{\phi}(i)$. 

%Similar to $\mathcal{SG}^{\phi}$, the Rabin acceptance condition for $\mathcal{M}$ is: $F^{\phi, \mathcal{C}_{d}, \mathcal{C}_{a}} = \{(L^{\phi, \mathcal{C}_{d}, \mathcal{C}_{a}}(i), K^{\phi, \mathcal{C}_{d}, \mathcal{C}_{a}}(i))\}_{i=1}^M$, with $(s,q,g_{d},g_{a}) \in L^{\phi, \mathcal{C}_{d}, \mathcal{C}_{a}}(i)$ iff $(s,q) \in L^{\phi}(i)$ and $(s,q,g_{d},g_{a}) \in K^{\phi,\mathcal{C}_{d}, \mathcal{C}_{a}}(i)$ iff $(s,q) \in K^{\phi}(i)$. 

A state  of $\mathcal{M}$ is $\mathfrak{s} := (s,q,g_{d},g_{a})$. 
A path on $\mathcal{M}$ is a sequence $\pi:=\mathfrak{s}_0 \mathfrak{s}_1 \dots$ such that $\mathbb{T}(\mathfrak{s}_{k+1}|\mathfrak{s}_k) > 0$, where $\mathbb{T}(\cdot)$ is the transition probability in $\mathcal{M}$. 
The path is accepting if it satisfies the Rabin acceptance condition. 
This corresponds to an execution in $\mathcal{SG}^{\phi}$ controlled by $\mathcal{C}_{d}$ and $\mathcal{C}_{a}$. 
%A probability space over $\mathcal{M}$ is defined in the usual way \cite{baier2008principles}. 
%
%\subsection{System Model} \label{Model}
%
%Consider a discrete-time finite-state system: 
%\begin{align}
%x(t+1)&=f(x(t),u_{d}(t),u_{a}(t), \mathtt{v}(t))
%\end{align}
%where $\mathtt{v}(t)$ represents a stochastic disturbance. 
%Notice that the transition from a state $x(t)$ to a state $x(t+1)$ is influenced by inputs from the defender ($u_{d}(t)$) and the adversary ($u_{a}(t)$). 
%This system can be abstracted as a stochastic game with finite state and action spaces using a simulation-based algorithm, similar to that in \cite{niu2018optimal}. 

To quantitatively reason about $\mathcal{M}$, we define a probability space following the treatment in \cite{baier2008principles}.  
The set of paths in $\mathcal{M}$, denoted $\mathtt{Paths}(\mathcal{M})$ forms the sample space. 
The set of events $\mathcal{F}$ is the smallest $\sigma-$algebra generated by \emph{cylinder sets} spanned by path fragments of finite length in $\mathcal{M}$. 
The cylinder set spanned by $\hat{\pi}:=\mathfrak{s}_0 \mathfrak{s}_1 \dots \mathfrak{s}_n$ is given by paths $\pi \in \mathtt{Paths}(\mathcal{M})$ that start with $\hat{\pi}$. 
This is denoted $Cyl(\mathfrak{s}_0 \mathfrak{s}_1 \dots \mathfrak{s}_n)$. 
Then, the (unique) probability measure on $\mathcal{F}$ for the events is given by $Pr^{\mathcal{M}}(Cyl(\mathfrak{s}_0 \mathfrak{s}_1 \dots \mathfrak{s}_n)):=\mathbb{P}(\mathfrak{s}_0)\bar{\mathbb{T}}(\mathfrak{s}_{1}|\mathfrak{s}_0) \bar{\mathbb{T}}(\mathfrak{s}_{2}|\mathfrak{s}_1) \dots \bar{\mathbb{T}}(\mathfrak{s}_{n}|\mathfrak{s}_{n-1})$. 
The probability of the LTL objective $\phi$ being satisfied in state $\mathfrak{s}$ is $Pr^{\mathcal{M}}(s \models \phi) := Pr^{\mathcal{M}}\{ \pi \in \mathtt{Paths}(\mathfrak{s}) | \pi \models \phi\}$. 
In the sequel, we write $\mathbb{P}(\mathcal{M} \models \phi):=\mathbb{P}(\mathcal{M}^{\phi, \mathcal{C}_{d}, \mathcal{C}_{a}}  \models \phi)$ to denote $Pr^{\mathcal{M}}(s_0 \models \phi)$. 
We direct the reader to Section 10.1 in \cite{baier2008principles} for a characterization of the probability spaces for different LTL objectives. 
\subsection{Problem Statement}\label{Stmt}

The goal is to synthesize a defender policy that maximizes the probability of satisfaction of an LTL specification under any adversary policy. 
Clearly, this will depend on the FSCs, $\mathcal{C}_{d}$ and $\mathcal{C}_{a}$. 
In this paper, we will assume that the size of the adversary FSC is fixed, and known to the defender. 
This can be interpreted as one way for the defender to have knowledge of the capabilities of an adversary. 
Future work will consider the problem for FSCs of arbitrary sizes. 
The problem is stated below.  
\begin{prob}\label{Prob}
Given a partially observable environment and an LTL formula, determine a defender policy specified by an FSC that maximizes the probability of satisfying the LTL formula under any adversary policy that is represented as an FSC of fixed size. \end{prob}
%
%Determine defender and adversary finite state controllers that result in a defender policy that maximizes the probability of a POSG satisfying an LTL specification, under any adversary policy. 
%That is, 
%\begin{align}
%\max_{\mathcal{C}_{d}}~ \min_{\mathcal{C}_{a}}~ \mathbb{P}(\mathcal{SG}^{\phi} \models \phi | \mathcal{C}_{d}, \mathcal{C}_{a}, |G_{a}| = G_A)
%\end{align}
%\end{prob}
%
%Optimizing over $\mathcal{C}_{d}$ and $\mathcal{C}_{a}$ indicates that the solution will depend on $|G_{d}|$, $\mu_{d}(\cdot)$, and $\mu_{a}(\cdot)$. 
%
%\section{Results}\label{Results}
%
%\subsection{LTL Satisfaction and Recurrent Sets}\label{LTLRecSet}
%
\section{LTL Satisfaction and Recurrent Sets}\label{LTLRecSet}

The first result in this section relates the probability of the LTL specification being satisfied by the product-POSG, denoted $\mathcal{SG}^{\phi} \models \phi$, in terms of recurrent sets of the GMC. 
We then present a procedure to generate recurrent sets of the GMC that additionally satisfy the LTL formula. 
The main result of this section relates Problem \ref{Prob} to determining FSCs that maximize the probability of reaching certain types of recurrent sets of the GMC. 
%The reader is referred to \cite{bhaskar2019finite} for the proofs of these results. 
%We conclude the section by presenting a procedure to determine the subset of states consistent with the Rabin acceptance conditions corresponding to the LTL formula that need to be visited infinitely often in steady state. 

Let $\mathcal{R}=\mathcal{R}^{\phi,\mathcal{C}_{d},\mathcal{C}_{a}}$ denote the recurrent states of $\mathcal{M}$ under FSCs $\mathcal{C}_{d}$ and $\mathcal{C}_{a}$. 
Let $\mathcal{R}^S:= (s,q)$ be the restriction of a recurrent state to a state of $\mathcal{SG}^{\phi}$. 
\begin{prop}\label{Proposn}%[\cite{bhaskar2019finite}]\label{Proposn}
$\mathbb{P}(\mathcal{M} \models \phi)=\mathbb{P}(\mathcal{SG}^{\phi} \models \phi| \mathcal{C}_{d}, \mathcal{C}_{a}) > 0$ if and only if there exists $\mathcal{C}_{d}$ such that for any $\mathcal{C}_{a}$, there exists a Rabin acceptance pair $(L^{\phi}(i),K^{\phi}(i))$ and an initial state of $\mathcal{M}$, $\bar{s}_0$, where the following conditions hold:
\begin{align}
&K^{\phi}(i) \cap \mathcal{R}^S \neq \emptyset \nonumber \\
\bar{s}_0&\rightarrow (K^{\phi}(i) \times G_{d} \times G_{a}) \cap \mathcal{R} \label{QualSat}\\
\bar{s}_0 &\centernot \rightarrow (L^{\phi}(i) \times G_{d} \times G_{a}) \cap \mathcal{R} \nonumber 
\end{align}
\end{prop} 
\begin{IEEEproof}
If for every $(L^{\phi}(i),K^{\phi}(i))$, at least one of the conditions in Equation (\ref{QualSat}) does not hold, then at least one of the following statements is true: 
\emph{i)}: no state that has to be visited infinitely often is recurrent; 
\emph{ii)}: there is no initial state from which a recurrent state that has to be visited infinitely often is accessible; 
\emph{iii)}: some state that has to be visited only finitely often in steady state is recurrent. 
This means $\mathcal{SG}^{\phi} \not \models \phi$ for all $\mathcal{C}_{d}$.

Conversely, if all the conditions in Equation (\ref{QualSat}) hold for some $(L^{\phi}(i),K^{\phi}(i))$, then $\mathcal{SG}^{\phi} \models \phi$ by construction.
\end{IEEEproof}

To quantify the satisfaction probability for a defender policy under any adversary policy, assume that the recurrent states of $\mathcal{M}$ are partitioned into recurrence classes $\{R_1,\dots,R_p\}$. 
This partition is maximal, in the sense that two recurrent classes cannot be combined to form a larger recurrent class, and all states within a given recurrent class communicate with each other \cite{sharan2014formal}. 
\begin{df}[$\phi-$feasible Recurrent Set]\label{PhiRecSet}
A recurrent set $R_k$ is \emph{$\phi-$feasible} under FSCs $\mathcal{C}_{d}$ and $\mathcal{C}_{a}$ if there exists $(L^{\phi}(i), K^{\phi}(i))$ such that $K^{\phi}(i) \cap R^S_k \neq \emptyset$ and $L^{\phi}(i) \cap R^S_k = \emptyset$.  
Let $\phi-RecSets^{\mathcal{C}_{d}, \mathcal{C}_{a}}$ denote the set of $\phi-$feasible recurrent sets under the respective FSCs. 
\end{df}

%Over infinite executions, a path of $\mathcal{M}$ will reach a recurrent set. 
%Let $\pi \rightarrow R$ denote this event. %the event that such a path will reach a recurrent set. 
Let $\pi \rightarrow R$ be the event that a path of $\mathcal{M}$ will reach a recurrent set.
Algorithm \ref{algoGenRecSets} returns $\phi-$feasible recurrent sets of $\mathcal{SG}^{\phi}$ under fixed FSCs $\mathcal{C}_{d}, \mathcal{C}_{a}$. 
\begin{algorithm}
\renewcommand{\algorithmiccomment}[1]{// #1}
\caption{Generate $\phi-$feasible Recurrent Sets for $\mathcal{SG}^{\phi}$ under FSCs $\mathcal{C}_{d}, \mathcal{C}_{a}$}\label{algoGenRecSets}
\begin{algorithmic}[1]
\REQUIRE $\mathcal{M}:=\mathcal{M}^{\phi,\mathcal{C}_{d}, \mathcal{C}_{a}}, \{L^{\phi}(i),K^{\phi}(i)\}_{i=1}^M$
\ENSURE $\{R_k\}$, that is recurrent and $\phi-$feasible
\STATE Induce digraph $\mathcal{G}$ of $\mathcal{M}$ of $\mathcal{SG}^{\phi}$ under $\mathcal{C}_{d}, \mathcal{C}_{a}$ as $(\mathfrak{S}, \mathcal{E})$, s.t. $\forall \mathfrak{s}_1, \mathfrak{s}_2 \in \mathfrak{S}: \mathfrak{s}_1 \rightarrow \mathfrak{s}_2 \in \mathcal{E} \Leftrightarrow \mathbb{T}(\mathfrak{s}_2|\mathfrak{s}_1) > 0$.
\STATE $\mathcal{C} = SCCs(\mathcal{G}) = \{C_1, \dots, C_N\}$ \COMMENT{\emph{strongly connected components of digraph}}
\STATE $RecSets:= \{R_1,\dots,R_p\}$ such that $R_i \in \mathcal{C}$ and $R_i$ is a \emph{sink SCC}
\STATE $\phi-RecSets^{\mathcal{C}_{d}, \mathcal{C}_{a}} = \emptyset$
\FOR {$j=1$ to $p$} 
\FOR {$i = 1$ to $M$} 
\IF {$(L^{\phi}(i) \cap R^S_j = \emptyset) \wedge (K^{\phi}(i) \cap R^S_j \neq \emptyset)$}
\STATE $\phi-RecSets^{\mathcal{C}_{d}, \mathcal{C}_{a}} = \phi-RecSets^{\mathcal{C}_{d}, \mathcal{C}_{a}} \cup R_j$
\ENDIF
\ENDFOR
\ENDFOR
\end{algorithmic}
\end{algorithm}

We have the following result:
%Theorem \ref{TheoremRecSet} states that 
%%determining defender FSCs that maximize the satisfaction probability for any adversary FSC as specified by 
%Problem \ref{Prob} is equivalent to determining $\mathcal{C}_{d}$ that maximizes the probability of reaching $\phi-$feasible recurrent sets of the GMC under any $\mathcal{C}_{a}$. 
%
\begin{thm}\label{TheoremRecSet} %[\cite{bhaskar2019finite}] \label{TheoremRecSet} 
The probability of satisfying an LTL formula $\phi$ in a POSG with policies $\mathcal{C}_{d}$ and $\mathcal{C}_{a}$ is equal to the probability of paths in the GMC (under the same FSCs) reaching $\phi-$feasible recurrent sets. That is,
\begin{align}\label{SatRecSet}
\mathbb{P}(\mathcal{SG}^{\phi} \models \phi | \mathcal{C}_{d}, \mathcal{C}_{a})&=\sum_{R \in \phi-RecSets^{\mathcal{C}_{d}, \mathcal{C}_{a}}}\mathbb{P} (\pi \rightarrow R)
\end{align}
\end{thm}
\begin{IEEEproof}
Since the recurrence classes are maximal, $\mathbb{P}(\pi \rightarrow (R_1 \cup \dots \cup R_p) ) = \sum_{k=1}^p \mathbb{P}(\pi \rightarrow R_k)$. 
From Definition \ref{PhiRecSet}, a $\phi-$feasible recurrent set will necessarily contain a Rabin acceptance pair. 
Therefore, the probability of $\mathcal{SG}^{\phi}$ satisfying the LTL formula under $\mathcal{C}_{d}$ and $\mathcal{C}_{a}$ is equivalent to the probability of paths on $\mathcal{M}$ leading to $\phi-$feasible recurrent sets, which is given by Equation (\ref{SatRecSet}).
%That is, $\mathbb{P}(\mathcal{SG}^{\phi} \models \phi | \mathcal{C}_{d}, \mathcal{C}_{a}) =  \sum_{R \in \phi-RecSets^{\mathcal{C}_{d}, \mathcal{C}_{a}}}\mathbb{P} (\pi \rightarrow R)$.} 
\end{IEEEproof}
\begin{cor}\label{CorRecSet}
From Theorem \ref{TheoremRecSet}, it follows that:
\begin{align}
\max_{\mathcal{C}_{d}} ~\min_{\mathcal{C}_{a}}~&\mathbb{P}(\mathcal{M} \models \phi)=\max_{\mathcal{C}_{d}} ~\min_{\mathcal{C}_{a}}~ \mathbb{P}(\mathcal{SG}^{\phi} \models \phi | \mathcal{C}_{d}, \mathcal{C}_{a})\nonumber \\&=\max_{\mathcal{C}_{d}} ~\min_{\mathcal{C}_{a}} \sum_{R \in \phi-RecSets^{\mathcal{C}_{d}, \mathcal{C}_{a}}}\mathbb{P} (\pi \rightarrow R)
\end{align}
\end{cor}

We note that Proposition \ref{Proposn}, Theorem \ref{TheoremRecSet}, and Corollary \ref{CorRecSet} address a broader class of problems than in Problem \ref{Prob} since they do not assume that the size of the adversary FSC is fixed. 
Corollary \ref{CorRecSet} also indicates that the objective of Problem \ref{Prob} can be formally expressed as:
\begin{align}
\max_{\mathcal{C}_{d}}~ \min_{\mathcal{C}_{a}}~ \mathbb{P}(\mathcal{M} \models \phi | |G_{a}| = G_A)
\end{align}
%
%Optimizing over $\mathcal{C}_{d}$ and $\mathcal{C}_{a}$ indicates that the solution will depend on $|G_{d}|$, $\mu_{d}(\cdot)$, and $\mu_{a}(\cdot)$. 
%
%\subsection{Value Iteration for POSGs}
\section{Determining Candidate FSCs of Fixed Sizes}\label{FindFSCs}
\begin{figure*}[t]
% ensure that we have normalsize text
\normalsize
% Store the current equation number.
%\setcounter{mytempeqncnt}{\value{equation}}
% Set the equation number to one less than the one
% desired for the first equation here.
% The value here will have to changed if equations
% are added or removed prior to the place these
% equations are referenced in the main text.
%\setcounter{equation}{5}
\begin{align}
O_{d}(o_{d}|s) O_{a}(o_{a}|s) \mu_{d}(g'_{d}, u_{d}|g_{d}, o_{d}) \mu_{a}(g'_{a}, u_{a}|g_{a}, o_{a}) \mathbb{T}^{\phi}((s',q')|(s,q),u_{d}, u_{a}) > 0 \label{BigEqn1}
\end{align}
\begin{align}
O_{d}(o_{d}|s) O_{a}(o_{a}|s) \mu_{d}(g''_{d}, u_{d}|g_{d}, o_{d}) \mu_{a}(g''_{a}, u_{a}|g_{a}, o_{a}) \mathbb{T}^{\phi}((s'',q'')|(s,q),u_{d}, u_{a}) > 0 \label{BigEqn2}
\end{align}
% Restore the current equation number.
%\setcounter{equation}{\value{mytempeqncnt}}
% IEEE uses as a separator
\hrulefill
% The spacer can be tweaked to stop underfull vboxes.
%\vspace*{2pt}
\end{figure*}
If the sizes of $\mathcal{C}_{d}$ and $\mathcal{C}_{a}$ are fixed, then their design is equivalent to determining the transition probabilities between their internal states. 
%We are guided by the treatment in \cite{sharan2014formal}. 
%However, our framework differs in that we additionally consider the effect of the presence of an adversary while aiming to satisfy an LTL specification. 
In this section, we present a heuristic procedure that uses only the most recent observations of the defender and adversary to generate a set of admissible FSC structures such that the resulting GMC will have a $\phi-$feasible recurrent set. 
%\textcolor{blue}{We use only the most recent observations of the defender and adversary. This results in the procedure having a computational complexity that is polynomial in the number of states of the GMC.} 
We show that the procedure has a computational complexity that is polynomial in the number of states of the GMC and additionally establish that this algorithm is sound.
\begin{df}
An algorithm is  \emph{sound} if any solution returned by it is the Boolean constant \emph{true} when evaluated on the output of the algorithm (i.e., every output is a correct output). 
It is \emph{complete} if it returns a result for any input, and reports `failure' if no solution exists. 
\end{df}

Let $\mathcal{I}_{*} : G_* \times \mathcal{O}_* \times G_* \times U_* \rightarrow \{0,1\}$, where $\mathcal{I}_*(g', u | g, o)$ $ = 1 \Leftrightarrow \mu_*(g',u|g,o) > 0$. 
$\mathcal{I}_*(\cdot)$ shows if an observation $o$ can enable the transition from an FSC state $g$ to $g'$ while issuing action $u$. 
We also assume that $\forall (g,o) \in G_* \times \mathcal{O}_*, \exists (g',u) \in G_* \times U_*$ such that  $\mathcal{I}_*(g', u | g, o) = 1$ \cite{sharan2014formal}. 
%Let a state in the GMC be denoted $\mathfrak{s}:= (s,q,g_{d},g_{a})$. 
%
%\begin{ass}\label{Nonblocking}
%$\mathcal{C}_{d}$ and $\mathcal{C}_{a}$ are \emph{nonblocking}. 
%That is, for every state in an FSC and every observation ($o_{d}$ or $o_{a}$, as the case may be), there exists a valid FSC state and a valid transition in the respective FSC (which together, will result in a transition in $\mathcal{SG}^{\phi}$). 
%\end{ass}
%
%This will preclude the adversary from setting all transition probabilities in $\mathcal{C}_{d}$ to zero. 
%
\begin{algorithm}[!h]
\caption{Generate candidate FSCs $\mathcal{C}_{d}, \mathcal{C}_{a}$}\label{algo1}
\begin{algorithmic}[1]
\REQUIRE $G_{d}$, $G_{a}$, $\mathcal{SG}^{\phi}$, $\mathcal{I}_{d}^o$, $\mathcal{I}_{a}^o$
\ENSURE Set of admissible FSC structures $\mathbb{I}:=(\mathbb{I}_{d}, \mathbb{I}_{a})$, %and transition probabilities, $(\mu_{d}(), \mu_{a}())$ 
such that GMC has a $\phi-$feasible recurrent set 
\STATE Induce digraph $\mathcal{G}$ of $\mathcal{M}$ of $\mathbb{SG}^{\phi}$ under $\mathcal{I}_{d}^o$ and $\mathcal{I}_{a}^o$ as $(\mathfrak{S}, \mathcal{E})$, s.t. $\forall \mathfrak{s}_1, \mathfrak{s}_2 \in \mathfrak{S}: \mathfrak{s}_1 \rightarrow \mathfrak{s}_2 \in \mathcal{E} \Leftrightarrow \mathbb{T}(\mathfrak{s}_2|\mathfrak{s}_1) > 0$. 
\STATE $\mathbb{I}_{d} = \mathbb{I}_{a} = \emptyset$
\STATE $\mathcal{C} = SCCs(\mathcal{G}) = \{C_1, \dots, C_N\}$
\FOR {$C \in \mathcal{C}$ \AND $(L^{\phi}(i),K^{\phi}(i)) \in F^{\phi}$}
\STATE $Bad_i=\{\mathfrak{s}' \notin C: \exists \mathfrak{s} \in C  \text{ s.t. } \mathfrak{s} \rightarrow \mathfrak{s}'\}$
\STATE $Bad_i= Bad_i \cup (C \cap (L^{\phi}(i) \times G_{d} \times G_{a}))$
\STATE $Good_i = C \cap (K^{\phi}(i) \times G_{d} \times G_{a})$
\STATE Set $\mathcal{I}_{*}(g'_{*}, u_*|g_*,o_*) = 1$ for all $g'_*, g_*, u_*, o_*$%, * \in \{d,a\}$
\WHILE {$\sum_{g'_*, u_*}\mathcal{I}_{*}(g'_{*}, u_*|g_*,o_*)  > 0 \forall o_*, g_*$ \AND $Bad_i \neq \emptyset$}
\STATE Choose $\mathfrak{s'}=(s',q',g_{d}',g_{a}') \in Bad_i$, \\$\mathfrak{s}''=(s'',q'',g_{d}'',g_{a}'') \in Good_i$
\FOR {$\mathfrak{s} = (s,q,g_{d},g_{a}) \in C \setminus Bad_i$}
\FOR {$u_{d} \in U_{d}$}
%\STATE $\mu_*(g'_*,u_*|\Phi_*,g_*,o_*) = \frac{\mathcal{I}_{*}(g'_{*}, u_*|g_*,o_*)}{\sum_{g'_*, u_*}\mathcal{I}_{*}(g'_{*}, u_*|g_*,o_*)}$
\STATE $\mu_*(g'_*,u_*|g_*,o_*) = \frac{\mathcal{I}_{*}(g'_{*}, u_*|g_*,o_*)}{\sum_{g'_*, u_*}\mathcal{I}_{*}(g'_{*}, u_*|g_*,o_*)}$
\IF {$\exists u_{a} \in U_{a}$ Eqn (\ref{BigEqn1}) holds }
\STATE $\mathcal{I}_{d}(g'_{d}, u_{d}|g_{d},o_{d}) \leftarrow 0$ \\$ \forall g'_{d}, g_{d} \in G_{d}$
\ENDIF
\ENDFOR
\FOR {$u_{a} \in U_{a}$}
%\STATE $\mu_*(g''_*,u_*|\Phi_*,g_*,o_*) = \frac{\mathcal{I}_{*}(g''_{*}, u_*|g_*,o_*)}{\sum_{g''_*, u_*}\mathcal{I}_{*}(g''_{*}, u_*|g_*,o_*)}$
\STATE $\mu_*(g''_*,u_*|g_*,o_*) = \frac{\mathcal{I}_{*}(g''_{*}, u_*|g_*,o_*)}{\sum_{g''_*, u_*}\mathcal{I}_{*}(g''_{*}, u_*|g_*,o_*)}$
\IF {$\forall u_{d} \in U_{d}$, Eqn (\ref{BigEqn2}) holds}
\STATE $\mathcal{I}_{a}(g''_{a}, u_{a}|g_{a},o_{a}) \leftarrow 0$%\forall g''_{a}, g_{a} \in G_{a}$
\ENDIF
\ENDFOR
\ENDFOR
\STATE $Bad_i = Bad_i \setminus \{\mathfrak{s}'\}$
\ENDWHILE
\STATE Construct digraph $\mathcal{G}_{new}$ of GMC of $\mathcal{SG}^{\phi}$ under modified $\mathcal{I}_{d}$ and $\mathcal{I}_{a}$
\STATE $\mathcal{C}_{new} = SCCs(\mathcal{G}_{new})$
\IF {$\exists \mathfrak{s} \in Good_i \text{ s.t. } \mathfrak{s}$ is \emph{recurrent} in $\mathcal{G}_{new}$}
\STATE $\mathbb{I} = (\mathbb{I}_{d} \cup \mathcal{I}_{d}, \mathbb{I}_{a} \cup \mathcal{I}_{a})$
\ENDIF
\ENDFOR
\end{algorithmic}
\end{algorithm}
 
In Algorithm \ref{algo1}, for defender and adversary FSCs with fixed number of states, we determine candidate $\mathcal{C}_{d}$ and $\mathcal{C}_{a}$ such that the resulting $\mathcal{M}$ will have a $\phi-$feasible recurrent set. 
We start with initial candidate structures $\mathcal{I}_*^o$ and induce the digraph of the resulting GMC (\emph{Line 1}). 
In our case, $\mathcal{I}_*^o$ is such that $\mathcal{I}^o_{*}(g'_*, u_*|g_*,o_*) = 1$ for all $g'_*, g_*, u_*, o_*$.  
%This MC might not contain a $\phi-$feasible recurrent set. 
We first determine the set of communicating classes of the GMC, which is equivalent to determining the strongly connected components (SCCs) of the induced digraph (\emph{Line 3}). 
A communicating class will be recurrent if it is a \emph{sink} SCC of the corresponding digraph. 
The states in $Bad_i$ are those in $C$ that are part of the Rabin accepting pair that has to be visited only finitely many times (and therefore, to be visited with very low probability in steady state) (\emph{Line 6}). 
$Bad_i$ further contains states that can be transitioned to from some state in $C$. 
This is because once the system transitions out of $C$, it will not be able to return to it in order to satisfy the Rabin acceptance condition (\emph{Line 5}) (and hence, $C$ will not be recurrent). 
$Good_i$ contains those states in $C$ that need to be visited infinitely often according to the Rabin acceptance condition (\emph{Line 7}). 

%Therefore, it is clear that in order to ensure Rabin acceptance, the states in $K^{\phi}(i) \times G_{d} \times G_{a}$ need to be recurrent in the global MC. 
The agents have access to a state only via their observations. 
%Therefore, if a state of $\mathcal{M}$ has to be made unreachable, we will disallow all corresponding transitions between internal states of the FSC corresponding to that observation. 
A defender action is forbidden if there exists an adversary action that will allow a transition to a state in $Bad_i$ under observations $o_{d}$ and $o_{a}$. 
This is achieved by setting corresponding entries in $\mathcal{I}_{d}$ to zero (\emph{Lines 12-17}). 
%To make a state in $Bad_i$ unreachable, we forbid defender actions under observation $o_{d}$ that lead to this state under corresponding adversary observation $o_{a}$ and some adversary action by setting corresponding entries in $\mathcal{I}_{d}$ to $0$. 
An adversary action is not useful if for every defender action, the probability of transitioning to a state in $Good_i$ is nonzero under $o_{d}$ and $o_{a}$. 
This is achieved by setting the corresponding entry in $\mathcal{I}_{a}$ to zero (\emph{Lines 18-23}). 
%
%Propositions \ref{Complexity} and \ref{Sound} indicate upper bounds on the computational complexity of Algorithm \ref{algo1} and guarantees on its soundness. 
%
\begin{prop}\label{Complexity}
Define $|\mathcal{O}| =|\mathcal{O}_{d}|+|\mathcal{O}_{a}|$ and $|U|=|U_{d}|+|U_{a}|$. 
Then, Algorithm \ref{algo1} has an overall computational complexity of $\mathbf{O}(|S|^2|G_{d}|^2|G_{a}|^2|\mathcal{O}||U|)$. 
\end{prop}
\begin{IEEEproof}
The overall complexity depends on: 
%\begin{itemize}
%\item 
\emph{(i)} Determining strongly connected components (\emph{Line 3}): 
This can be done in $\mathbf{O}(|\mathfrak{S}|+|\mathcal{E}|)$ \cite{tarjan1972depth}. 
Since $|\mathfrak{S}| = |S||G_{d}||G_{a}|$ and $|\mathcal{E}| \leq |\mathfrak{S}|^2$, this is $\mathbf{O}(|S|^2|G_{d}|^2|G_{a}|^2)$ in the worst case, and  
%\item 
\emph{(ii)} Determining the structures in \emph{Lines 9-26}: 
This is $\mathbf{O}(|\mathfrak{S}|(|\mathcal{O}_{d}|+|\mathcal{O}_{a}|)(|\mathfrak{S}|(|U_{d}|+|U_{a}|))$. 
%\end{itemize}
The result follows by combining the two terms. 
\end{IEEEproof}
%
%Moreover, under the current defender and adversary observations, if for some adversary action, every defender action leads to a state in $Good_i$, then the corresponding entry in $\mathcal{I}_{a}$ is set to $0$ (\emph{Line 9-25}). 
%
%If the modified global MC under $\mathcal{C}_{d}$ and $\mathcal{C}_{a}$ determined by the structure after eliminating the `bad' states has a `good' state that is also recurrent, then this state is $\phi-$feasible, and the resulting structure is a feasible structure. 
\begin{prop}\label{Sound}
Algorithm \ref{algo1} is sound. 
%That is, each feasible FSC structure $(\mathcal{I}_{d},\mathcal{I}_{a})$ in $\mathbb{I}$ will have at least one $\phi-$feasible recurrent set. 
\end{prop}
\begin{IEEEproof}
This is by construction. 
The output of the algorithm is a set $\{\mathcal{I}_{d}^i,\mathcal{I}_{a}^i\}_{i=1}^W$ such that the resulting GMC for each case has a state that is recurrent and has to be visited infinitely often. 
This state, by Definition \ref{PhiRecSet}, belongs to $\phi-RecSet^{\mathcal{C}_{d}^i,\mathcal{C}_{a}^i}$. 
Moreover, if the algorithm returns a nonempty solution, a solution to Problem \ref{Prob} will exist since the FSCs are proper. 
\end{IEEEproof}

Algorithm \ref{algo1} is \emph{suboptimal} since we only consider the most recent observations of the defender and adversary. 
It is also not complete, since there might be a feasible solution that cannot be determined by the algorithm. 
If no FSC structures of a particular size is returned by Algorithm \ref{algo1}, a heuristic is to increase the number of states in the defender FSC by one, and run the Algorithm again. 
Once we obtain proper FSC structures of fixed sizes, we will show in Section \ref{FSCVary} that the satisfaction probability can be improved by adding states to the defender FSC in a principled manner (for adversary FSCs of fixed size). Algorithm \ref{algo1} and Proposition \ref{Sound} will allow us to restrict our treatment to proper FSCs for the rest of the paper.
%However, since the algorithm, by construction, returns FSC structures such that the resulting global MC has a $\phi-$feasible recurrent set, the maximum satisfaction probability of LTL satisfaction under any adversary policy exists. 
%
%\begin{rem}
%For $\mathcal{C}_{d}$ and $\mathcal{C}_{a}$ of fixed sizes and structures $\mathcal{I}_{d}$ and $\mathcal{I}_{a}$, a solution to Problem \ref{Prob} is:
%\begin{align}
%\max_{\Phi_{d}}~ \min_{\Phi_{a}} ~\mathbb{P}(\mathcal{SG}^{\phi} \models \phi | \mathcal{C}_{d}, \mathcal{C}_{a})
%\end{align}
%\end{rem}
%
%This follows from the fact that for fixed FSC sizes and structures, the properties of a set (recurrent or transient) in the GMC will not change. 
%What remains then is to choose the transition probabilities appropriately. 
%For a softmax parameterization, this computation is presented in \cite{sharan2014formal}, and we omit it for want of space. 
%
\section{Value Iteration for POSGs}\label{POSGValIter}

In this section, we present a value-iteration based procedure to maximize the probability of satisfying the LTL formula $\phi$ for FSCs $\mathcal{C}_{d}$ and $\mathcal{C}_{a}$ of fixed sizes. 
We prove that the procedure converges to a unique optimal value, corresponding to the Stackelberg equilibrium. 
%Hence, we can extract the optimal control policy using the proposed VI procedure.

Notice that in Equation (\ref{TransFn}), the defender and adversary policies are specified as probability distributions over the next FSC internal state and the respective agent action, and conditioned on the current FSC internal state and the agent observation. 
With $*\in \{d,a\}$, we rewrite these in terms of a mapping $\hat{\mu}_*: G_* \times S \times G_* \times U_* \rightarrow [0,1]$:  
\begin{align}
\hat{\mu}_{*}(g'_*,u_*|g_*,s):=\sum_{o_* \in \mathcal{O}_*}O_*(o_*|s)\mu_*(g'_*,u_*|g_*,o_*)
\end{align}

This will allow us to express Equation (\ref{TransFn}) as:  
\begin{align}
&Pr^{\phi}((s',q'),g'_{d},g'_{a}|(s,q),g_{d},g_{a}) \nonumber \\
&=\sum_{u_{d}}\sum_{u_{a}}\hat{\mu}_{d}(g'_{d},u_{d}|g_{d},s)\hat{\mu}_{a}(g'_{a},u_{a}|g_{a},s)\nonumber \\ &\quad \quad \quad \quad \quad \mathbb{T}^{\phi}((s',q'),g'_{d},g'_{a}|(s,q),g_{d},g_{a})\label{ModTransProb}
\end{align}

Define a value $V$ over the state space of the GMC representing the probability of satisfying the LTL formula $\phi$ when starting from a state of the GMC. 
Additionally, define and characterize the following operators:
\begin{align*}
(T_{\hat{\mu}_{d}\hat{\mu}_{a}}V)(\mathfrak{s})&=\sum_{\mathfrak{s}'}Pr(\mathfrak{s}'|\mathfrak{s}) V(\mathfrak{s}'); \\
(T_{\hat{\mu}_{d}}V)(\mathfrak{s})&=\min_{\hat{\mu}_{a}}\sum_{\mathfrak{s}'}Pr(\mathfrak{s}'|\mathfrak{s}) V(\mathfrak{s}'); \\
(TV)(\mathfrak{s})&=\max_{\hat{\mu}_{d}}~\min_{\hat{\mu}_{a}}~\sum_{\mathfrak{s}'}Pr(\mathfrak{s}'|\mathfrak{s}) V(\mathfrak{s}')
\end{align*}
where $Pr(\mathfrak{s}'|\mathfrak{s})$ is the transition probability in the GMC induced by policies $\hat{\mu}_{d}$ and $\hat{\mu}_{a}$ (Equation (\ref{ModTransProb})).  
%We have the following result: 
%
\begin{prop}\label{ValuePOSG}
Let 
\begin{align}
V((s,q),&g_{d},g_{a})\nonumber \\&=\max_{\hat{\mu}_{d}}~\min_{\hat{\mu}_{a}}~Pr(\phi|((s,q),g_{d},g_{a})).\label{SatProb}
\end{align}
Then
\begin{multline}\label{eq:value vector}
    V((s,q),g_{d},g_{a})=\\\max_{\hat{\mu}_{d}}~\min_{\hat{\mu}_{a}}\sum_{((s',q'),g'_{d},g'_{a})}\sum_{u_{d}}\sum_{u_{a}}\Bigg(\hat{\mu}_{d}(g'_{d},u_{d}|g_{d},s)\\\times \hat{\mu}_{a}(g'_{a},u_{a}|g_{a},s)\\ \times
    T^\phi((s',q')|(s,q),u_{d},u_{a})V((s',q'),g'_{d},g'_{a}))\Bigg)
\end{multline}
Conversely, if the value vector $V$ satisfies Equation \eqref{eq:value vector}, then Equation (\ref{SatProb}) holds true. 
Moreover, $V$ is unique.
\end{prop}

Before proving Proposition \ref{ValuePOSG}, we will need some intermediate results. 
Inequalities in the proofs of these statements are true element-wise.
\begin{thm}\label{MCT}[Monotone Convergence Theorem]\cite{royden2010real}
If a sequence is monotone increasing and bounded from above, then it is a convergent sequence. 
\end{thm}
\begin{lm}\label{lemma:DP operator}
Let $V$ be the satisfaction probability obtained under any pair of policies $\hat{\mu}_{d}$ and $\hat{\mu}_{a}$, where $\hat{\mu}_{a}$ is the best response to $\hat{\mu}_{d}$. Let $T^k$ be the operation that composes the $T$ operator $k$ times, and $V^k$ be the corresponding value obtained (i.e., $T^kV:=V^k$). %when composing the $T$ operator $k$ times. 
Then, there exists a value $V^*$ such that  
%\begin{equation*}
    $\lim_{k\rightarrow\infty}T^kV=V^*$.
%\end{equation*}
\end{lm}
\begin{IEEEproof}
We show Lemma \ref{lemma:DP operator} by showing that the sequence $V^k=T^kV$ is bounded and monotone. 

We first show boundedness. By definition of the operator $T$, $V^{k+1}$ is obtained as a convex combination of $V^k$. 
Since $V$ is the satisfaction probability, it is in $[0,1]$. 
Thus, $V^0$ is bounded, and consequently, $T^kV$ is bounded for all $k$.

We next show monotonicity by induction. 
We have that $V^0$ is the value function associated with a control policy $\hat{\mu}_{d}$. 
Denote the best response of the adversary to $\hat{\mu}_{d}$ as $\hat{\mu}_{a}$. 
Let $V^1:=TV^0$. % be the value function obtained by applying operator $T$ on $V^0$. 
From the definitions of $T$ and $T_{\hat{\mu}_{d}}$, we have $TV^0\geq T_{\hat{\mu}_{d}}V^0$. 
%Using the fact that $\hat{\mu}_{a}$ is the best response to $\hat{\mu}_{d}$, 
Furthermore, $V^0=T_{\hat{\mu}_{d}}V^0$ since
\begin{align*}
&T_{\hat{\mu}_{d}}V^0(\mathfrak{s})=\min_{\hat{\mu}'_{a}}\sum_{\mathfrak{s}'} \sum_{u_{d}} \sum_{u_{a}}\Bigg(V^k(\mathfrak{s}') \hat{\mu}_{d}(g'_{d},u_{d}|g_{d},s)\\&\quad \quad \times\hat{\mu}_{a}(g'_{a},u_{a}|g_{a},s) \mathbb{T}^{\phi}((s',q')|(s,q),u_{d},u_{a})\Bigg)=V^0
\end{align*}
by the definition that $\hat{\mu}'_{a}$ is the best response of $\hat{\mu}_{d}$. 
%Furthermore, we have that $V^0=T_{\hat{\mu}_{d}}V^0$ since the operator $T_{\hat{\mu}_{d}}$ models the best response from the adversary, i.e., applying operator $T_{\hat{\mu}_{d}}$ is equivalent to computing $\hat{\mu}_{a}$ \footnote{D.P. Bertsekas, \emph{Dynamic Programming and Optimal Control}, Athena Scientific Belmont, MA, 1995, vol.1, no.2.}. 
Therefore, we have that $V^1=TV^0\geq T_{\hat{\mu}_{d}}V^0=V^0$. This gives us $V^1\geq V^0$, which serves as the base case for the induction. 
Consider iteration $k$. Suppose $T^{k-1}V\leq T^kV$. We then show $T^kV \leq T^{k+1}V$. We have: 
\begin{align*}
T^{k+1}V&\geq \min_{\hat{\mu}_{a}}\sum_{\mathfrak{s}'}Pr(\mathfrak{s}'|\mathfrak{s})V^{k}(\mathfrak{s}')
\geq\min_{\hat{\mu}_{a}}\sum_{\mathfrak{s}'}Pr(\mathfrak{s}'|\mathfrak{s}) V^{k-1}(\mathfrak{s}')
=V^k.
\end{align*}
The first inequality holds by the definition of $T$, the second inequality holds by the induction hypothesis that $V^k\geq V^{k-1}$, and the last equality holds by construction of a control policy.
The existence of $V^*$ such that $\lim_{k\rightarrow\infty}T^kV=V^*$ follows from Theorem \ref{MCT}.
%Then we show that the value function $V^1=TV^0$ obtained by applying operator $T$ on $V^0$ satisfies $V^1\geq V^0$, which will further server as our induction base. Consider the set of states $\mathfrak{s}$ in $\phi$-feasible recurrent sets. We have that $V^1(\mathfrak{s})=V^0(\mathfrak{s})=1$. The reason is that once the $\phi$-feasible recurrent set is reached, we can always find a control policy to remain in the $\phi$-feasible recurrent set. Then we consider the set of states $\mathfrak{s}$ not in $\mathfrak{s}$ in $\phi$-feasible recurrent sets. If $V^1(\mathfrak{s})\geq V^0(\mathfrak{s})$, then we can conclude $V^1\geq V^0$. If there exists some state $\mathfrak{s}$ such that $V^1(\mathfrak{s})< V^0(\mathfrak{s})$. Then let $\hat{\mu}^1_{d}(\mathfrak{s})=\hat{\mu}^0_{d}(\mathfrak{s})$, we can achieve value function $V^0(\mathfrak{s})>V^1(\mathfrak{s})$, which implies that $V^1$ does not satisfy $V^1=TV^0$ and hence contradicts our definition of $V^1$. Therefore, we have that $V^1\geq V^0$. 
%
%
%Finally we have $\lim_{k\rightarrow\infty}T^kV=V^*$ by Theorem \ref{MCT}. 
\end{IEEEproof}

We now prove Proposition \ref{ValuePOSG}. 
%The proof is along the lines of the proof of Lemma 1 in \cite{niu2018optimal}. 
%
\begin{IEEEproof}
We first prove the forward direction by contradiction, i.e., if Equation (\ref{SatProb}) holds then Equation (\ref{eq:value vector}) holds. Suppose $\hat{\mu}_{d}$ is a Stackelberg equilibrium policy with satisfaction probability being $V$, while Equation \eqref{eq:value vector} does not hold. Since $\hat{\mu}_{d}$ is the Stackelberg equilibrium policy, $V=T_{\hat{\mu}_{d}}V$. 
This is because, given $\hat{\mu}_{d}$, the stochastic game is an MDP, for which $V$ is the optimal value \cite{bertsekas2015dynamic}. 
From the definitions of $T$ and $T_{\hat{\mu}_{d}}$, we must have $T_{\hat{\mu}_{d}}V\leq TV$. 
Composing the operators $k$ times and letting $k\rightarrow\infty$,
\begin{equation*}
V=\lim_{k\rightarrow\infty}T^k_{\hat{\mu}_{d}}V\leq \lim_{k\rightarrow\infty}T^kV=V^*,
\end{equation*}
where the first equality holds by the assumption that $V$ is the satisfaction probability and $\hat{\mu}_{d}$ is the Stackelberg equilibrium policy, the last equality holds by Lemma \ref{lemma:DP operator}. %Therefore, $V\leq V^*$. %, where $V^*$ is the limit of the sequence of values obtained by composing operator $T$. 
If $V=V^*$, Eqn. (\ref{eq:value vector}) is satisfied, which contradicts our assumption that Eqn. (\ref{SatProb}) holds while (\ref{eq:value vector}) does not hold. If $V\neq V^*$, then there must exist a state $\mathfrak{s}$ such that $V(\mathfrak{s}) < V^*(\mathfrak{s})$. 
This means that there is a policy (different from $\hat{\mu}_{d}$) corresponding to $V^*$ for which we achieve a higher satisfaction probability starting at state $\mathfrak{s}$. 
This violates our assumption that $\hat{\mu}_{d}$ is the equilibrium policy. We must then have that Eqn. (\ref{eq:value vector}) holds given that Eqn. (\ref{SatProb}) holds.
%
%We next prove the backward direction, i.e., Eqn. (\ref{SatProb}) holds given that Eqn. (\ref{eq:value vector}) holds. Suppose $\hat{\mu}_{d}$ is control policy that satisfies Eqn. (\ref{eq:value vector}) with $V$ being the corresponding value function, while Eqn. (\ref{SatProb}) does not hold. Let $\mu'$ be the SE control policy and $V'$ be the corresponding SE satisfaction probability. As we have shown earlier, $V'$ solves Eqn. \eqref{eq:value vector} since $V'$ satisfies Eqn. \eqref{SatProb}, i.e., $V'=\lim_{k\rightarrow\infty}T^kV'$. Then by our assumption that $V'$ is the value associated with SE control policy $\mu'$, we have that $$V'=\lim_{k\rightarrow\infty}T^kV'\geq \lim_{k\rightarrow\infty}T^k_{\hat{\mu}_{d}}V.$$ Given that $\hat{\mu}_{d}$ solves Eqn. \eqref{eq:value vector}, we have that $T_{\hat{\mu}_{d}}V=TV$. Therefore, we have that $$V'=\lim_{k\rightarrow\infty}T^kV'\geq \lim_{k\rightarrow\infty}T^k_{\hat{\mu}_{d}}V=V.$$ control policy $\hat{\mu}_{d}$ is better than the SE control policy $\mu'$, which violates our assumption that $\mu'$ is the SE control policy. IF $V'=V$, then the assumption that $V'$ does not satisfy Eqn. \eqref{eq:value vector} is violated. If $V'>V$, then control policy $\hat{\mu}_{d}$ is not the solution to Eqn. \eqref{eq:value vector} since control policy $\mu'$ is better. Hence, we must have Eqn. (\ref{SatProb}) holds given that Eqn. (\ref{eq:value vector}) holds.

We next prove uniqueness of the $V$. 
Let $\hat{V}$ and $V$ be two solutions to Eqn. \eqref{eq:value vector}, and let $\hat{\mu}_{d}$ and $\mu_{d}$ denote the corresponding control policies. From the definitions of $T$ and $T_{\hat{\mu}_{d}}$, we have that $V=TV\geq T_{\hat{\mu}_{d}}V$. Composing the operators on both sides $k$ times and letting $k\rightarrow\infty$,
	\begin{equation*}
	V=\lim_{k\rightarrow\infty}T^kV\geq \lim_{k\rightarrow\infty}T^k_{\hat{\mu}_{d}}V=\hat{V},
	\end{equation*}
where the first equality holds by the assumption that $\hat{\mu}_{d}$ is the equilribrium policy, and the second equality holds by the fact that $\hat{V}$ is the unique fixed point of operator $T_{\hat{\mu}_{d}}$ (Proposition $2.2.1$ in Volume $2$ of \cite{bertsekas2015dynamic}). %\footnote{Corollary 2.1.1 in D.P. Bertsekas, \emph{Dynamic Programming and Optimal Control}, Athena Scientific Belmont, MA, 1995, vol.1, no.2.}. 
Following a similar argument as before, we have the following inequality:
 \begin{equation*}
 \hat{V}=\lim_{k\rightarrow\infty}T^k\hat{V}\geq \lim_{k\rightarrow\infty}T^k_{\hat{\mu}_{d}}\hat{V}=V.
 \end{equation*}
We have that both $V\geq \hat{V}$ and $\hat{V}\geq V$ are true, which gives us $V=\hat{V}$. This implies that the value $V$ is unique.

We finally show that if  Eqn. (\ref{eq:value vector}) holds then Eqn. (\ref{SatProb}) holds. 
We observe that the value function at equilibrium %\footnote{Every two person finite game admits a Stackelberg strategy for the leader.} 
satisfies \eqref{eq:value vector}, and the solution is unique. Therefore, any solution to \eqref{eq:value vector} must be a Stackelberg equilibrium \cite{conitzer2016stackelberg}.
%We finally show that if  Eqn. (\ref{eq:value vector}) holds then Eqn. (\ref{SatProb}) holds. Suppose $\hat{\mu}_{d}$ is control policy that satisfies Eqn. (\ref{eq:value vector}) with $V$ being the corresponding value function, while Eqn. (\ref{SatProb}) does not hold. Let $\mu'$ be the SE control policy and $V'$ be the corresponding SE satisfaction probability. As we have shown earlier, $V'$ solves Eqn. \eqref{eq:value vector} since $V'$ satisfies Eqn. \eqref{SatProb}, i.e., $V'=\lim_{k\rightarrow\infty}T^kV'$. Then by our assumption that $V'$ is the value associated with SE control policy $\mu'$, we have that $$V'=\lim_{k\rightarrow\infty}T^kV'\geq \lim_{k\rightarrow\infty}T^k_{\hat{\mu}_{d}}V.$$ Given that $\hat{\mu}_{d}$ solves Eqn. \eqref{eq:value vector}, we have that $T_{\hat{\mu}_{d}}V=TV$. Therefore, we have that $$V'=\lim_{k\rightarrow\infty}T^kV'\geq \lim_{k\rightarrow\infty}T^k_{\hat{\mu}_{d}}V=V.$$ control policy $\hat{\mu}_{d}$ is better than the SE control policy $\mu'$, which violates our assumption that $\mu'$ is the SE control policy. IF $V'=V$, then the assumption that $V'$ does not satisfy Eqn. \eqref{eq:value vector} is violated. If $V'>V$, then control policy $\hat{\mu}_{d}$ is not the solution to Eqn. \eqref{eq:value vector} since control policy $\mu'$ is better. Hence, we must have Eqn. (\ref{SatProb}) holds given that Eqn. (\ref{eq:value vector}) holds.
\end{IEEEproof}

Proposition \ref{ValuePOSG} indicates that value-iteration algorithms can be used to determine optimal policies $\hat{\mu}_{d}$ and $\hat{\mu}_{a}$. 
Given a policy $\hat{\mu}_{d}$ and observation function $O_{d}$, we will be able to compute $\mu_d$ by solving a system of linear equations. 

A value-iteration based procedure to solve the POSG under an LTL specification is proposed in Algorithm \ref{algo1a}. 
The value $V(\mathfrak{s})$ is greedily updated at every iteration by computing the policy according to Proposition \ref{ValuePOSG}. 
The algorithm terminates when the difference in $V(\cdot)$ in consecutive iterations is below a pre-specified threshold. 
\begin{algorithm}
\renewcommand{\algorithmiccomment}[1]{// #1}
\caption{Maximizing probability of satisfying LTL formula $\phi$ under fixed FSCs $\mathcal{C}_{d}, \mathcal{C}_{a}$}\label{algo1a}
\begin{algorithmic}[1]
\REQUIRE $\mathcal{M}:=\mathcal{M}^{\phi,\mathcal{C}_{d}, \mathcal{C}_{a}}, \{L^{\phi}(i),K^{\phi}(i)\}_{i=1}^M, \epsilon$ (threshold)
\ENSURE $V \in \mathbb{R}^{|S| \times |Q| \times |G_{d}| \times |G_{a}|}$
\STATE Determine $\phi-RecSets^{\mathcal{C}_{d}, \mathcal{C}_{a}}$ using Algorithm \ref{algoGenRecSets}
\STATE $\hat{\mu}_{*}(g'_*,u_*|g_*,s):=\sum \limits_{o_*}O_*(o_*|s)\mu(g'_*,u_*|g_*,o_*)$, $*\in \{d,a\}$
\STATE $V^0(\mathfrak{s}) \leftarrow 0$
\STATE $V^1(\mathfrak{s}) \leftarrow 1$ if %$(s,q) \in \mathcal{R}^S$, 
$\mathfrak{s} \in \phi-RecSets^{\mathcal{C}_{d}, \mathcal{C}_{a}}$, 
and $V^1(\mathfrak{s}) \leftarrow 0$, else
\STATE $k \leftarrow 0$
\WHILE {$\max \limits_{\mathfrak{s}}\{V^{k+1}(\mathfrak{s})-V^k(\mathfrak{s})\} > \epsilon$}
\STATE $k \leftarrow k+1$
\FOR {$\mathfrak{s} \notin \phi-RecSets^{\mathcal{C}_{d}, \mathcal{C}_{a}}$} 
\STATE $V^{k+1}(\mathfrak{s}) \leftarrow \max \limits_{\hat{\mu}_{d}}\min \limits_{\hat{\mu}_{a}} \sum \limits_{\mathfrak{s}'} \sum \limits_{u_{d}} \sum \limits_{u_{a}}\Bigg(V^k(\mathfrak{s}')% 
\hat{\mu}_{d}(g'_{d},u_{d}|g_{d},s)$\\$\quad \quad \times\hat{\mu}_{a}(g'_{a},u_{a}|g_{a},s)%$\\$\quad \quad \times 
\mathbb{T}^{\phi}((s',q')|(s,q),u_{d},u_{a})\Bigg)$
\ENDFOR
\ENDWHILE
\RETURN $V (=V^k(\mathfrak{s}))$
\end{algorithmic}
\end{algorithm}
\begin{prop}
For any $\epsilon > 0$, there exist $K, V$ %$\delta, K, V$ 
such that $||V^k(\mathfrak{s})-V||_{\infty} < \epsilon$ for all $k>K$. 
Further, $V$ satisfies the value in Proposition \ref{ValuePOSG} and is within the $\epsilon$-neighborhood of the value function at Stackelberg equilibrium. 
\end{prop}
\begin{IEEEproof}
Notice that $V^1(\mathfrak{s}) \geq V^0(\mathfrak{s})$. 
Since $V^1(\mathfrak{s}) = 0$ for %$(s,q) \notin \mathcal{R}^S$, 
$(s,q) \notin \phi-RecSets^{\mathcal{C}_{d}, \mathcal{C}_{a}}$, 
$V^2(\mathfrak{s}) \geq V^1(\mathfrak{s})$. 
We induct on $k$.%to show monotonicity. 
Let $\hat{\mu}_{d}^k$ be the optimal defender policy at step $k$. 
Then, 
\begin{align*}
V^{k+1}(\mathfrak{s}) &\geq \min_{\hat{\mu}_{a}}\sum_{\mathfrak{s}'} \sum_{u_{d}} \sum_{u_{a}}\Bigg(V^k(\mathfrak{s}') \hat{\mu}_{d}(g'_{d},u_{d}|g_{d},s)\\&\quad \quad \times\hat{\mu}_{a}(g'_{a},u_{a}|g_{a},s) \mathbb{T}^{\phi}((s',q')|(s,q),u_{d},u_{a})\Bigg)\\
&\geq \min_{\hat{\mu}_{a}}\sum_{\mathfrak{s}'} \sum_{u_{d}} \sum_{u_{a}}\Bigg(V^{k-1}(\mathfrak{s}') \hat{\mu}_{d}(g'_{d},u_{d}|g_{d},s)\\&\quad  \times\hat{\mu}_{a}(g'_{a},u_{a}|g_{a},s) \mathbb{T}^{\phi}((s',q')|(s,q),u_{d},u_{a})\Bigg)=V^k(\mathfrak{s})
\end{align*}
The first inequality holds because $V^{k+1}(\mathfrak{s})$ is the value obtained by the maximizing policy, and dominates the value achieved by any other policy. 
The second and last inequalities follow from the induction hypothesis and definition of $V^k(\mathfrak{s})$ respectively. 
%The second inequality follows from the induction hypothesis, and the last equality follows from the definition of $V^k(\mathfrak{s})$. 
%
Further, for each state, $V^k(\mathfrak{s})$ is bounded since it is a convex combination of terms that are $\leq 1$. 
Let $V$ be the set of limit points so that $K$ can be chosen such that $||V^k(\mathfrak{s}) - V||_{\infty} < \epsilon$ for $k>K$. 

Since $V^k(\mathfrak{s})$ converges, it is a Cauchy sequence. 
Therefore, for every $\epsilon > 0$, there exists $K$ sufficiently large, such that for all $k > K$, $|V^k(\mathfrak{s}) - V^{k+1}(\mathfrak{s})| < \epsilon$. 
From \emph{Line 8} of Algorithm \ref{algo1a}, this shows that $V$ is within an $\epsilon-$neighborhood of a Stackelberg equilibrium for every $\epsilon > 0$. 
\end{IEEEproof}

%\textcolor{blue}{Additional guarantees on the convergence of Algorithm \ref{algo1a} can be established. 
%along the lines of Proposition 4 of \cite{niu2018optimal}. 
A minor modification of the value update gives the following result on the termination of Algorithm \ref{algo1a} \cite{niu2018optimal}.% to conclude this section.} 
\begin{prop}[\cite{niu2018optimal}, Proposition 4]
Suppose that at time $k$, the value update in \emph{Line 8} of Algorithm \ref{algo1a} is performed as shown if the right hand side term is greater than $(1+\epsilon)V^k(\mathfrak{s})$, and $V^{k+1}(\mathfrak{s}) = V^k(\mathfrak{s})$ otherwise. 
Then, Algorithm \ref{algo1a} converges to a value $V$ that satisfies $||V^{k+1}(\mathfrak{s}) - V^k(\mathfrak{s})||_{\infty} < \epsilon$ in at most $N^*$ iterations, where $N^* = |S| |Q| |G_{d}| |G_{a}| \max \limits_{\mathfrak{s}} \Big\{\log(\frac{1}{V_{min}(\mathfrak{s})})/\log(1+\epsilon)\Big\}$. Here $V_{min}(\mathfrak{s})$ is the smallest non-zero value of $V^k(\mathfrak{s})$.
\end{prop}
\section{Adding States to $\mathcal{C}_{d}$}\label{FSCVary}

Whenever Algorithm \ref{algo1} returns a non-empty solution, the FSCs are proper (Definition \ref{FSCProper}). In this case, there is  a nonzero probability of visiting a state in a $\phi-$feasible recurrent set of $\mathcal{SG}^{\phi}$ under FSCs $\mathcal{C}_{d}, \mathcal{C}_{a}$. 

It might be the case that the probability of satisfying the LTL formula can be increased by adding states to $\mathcal{C}_{d}$. 
In doing so, it is important to ensure that adding states to $\mathcal{C}_{d}$ does not decrease this satisfaction probability when compared to the satisfaction probability for $\mathcal{C}_{d}$ with fewer states. 
This lends itself to a policy iteration like approach. 
Policy iteration \cite{bertsekas2015dynamic} is a procedure that alternates between policy evaluation and policy improvement until convergence to a Stackelberg equilibrium. 
The policy evaluation step involves solving a Bellman equation, while the policy improvement step then `greedily' chooses a policy that maximizes the satisfaction probability. 

In this section, we will assume that the size of $\mathcal{C}_{a}$ is fixed. 
Intuititively, this means that the defender is aware of the capabilities of the adversary. % which is reflected by the size of $\mathcal{C}_{a}$, i.e., $|G_{a}|=G_A$. 
However, the defender does not know the transition probabilities in $\mathcal{C}_{a}$, which means it needs to determine the transition probabilities of its own FSC in order to be robust against the worst-case transition probabilities between states in $\mathcal{C}_{a}$. 
Future work will seek to address the situations when the defender knows the nature of the strongest possible adversary (this will correspond to $|G_{a}| \leq G_A$), and when the defender does not know anything about the abilities of an adversary. 

We will work with the `value' of a state $g_{d} \in G_{d}$ in $\mathcal{C}_{d}$. 
Denote this by $V_{g_{d}}(\mathfrak{s})$. 
From the defender's perspective, the transition probabilities in $\mathcal{C}_{d}$ are influenced by its \emph{belief} of the state of $\mathcal{SG}^{\phi}$ under $\mathcal{C}_{d}$ and $\mathcal{C}_{a}$. 
The \emph{belief} is a (prior) probability distribution over the states of the POSG. 
Then, the value of $g_{d} \in G_{d}$ under belief $b = \{b_1,\dots,b_{|S|}\}$ can be written as $V_{g_{d}}(b) = \sum_i b_i V_{g_{d}}(s_i)$. 
The value function for a belief $b$ is then given by
\begin{align}
V(b)&:= \max_{g_{d}}~ V_{g_{d}}(b)\label{ValForBelief}
\end{align} 
\begin{figure}[!h]
 \centering
  \includegraphics[width=2.65 in]{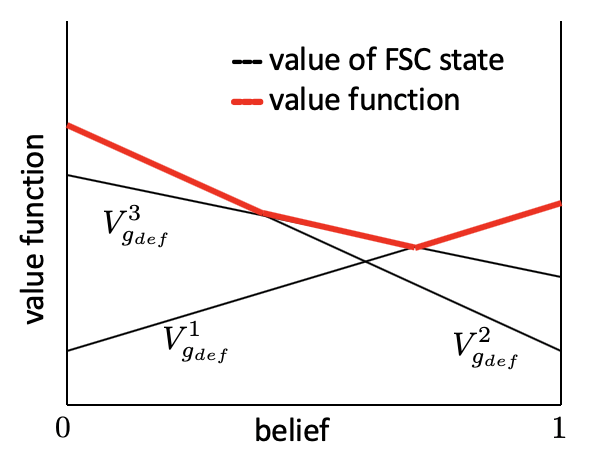} 
\caption{Value function of a two-state POSG with three states in $\mathcal{C}_{d}$. The value of each FSC state is linear in the belief (black lines). The value function is the point-wise maximum of the values of the FSC states (red curve).}\label{ValFuncFig}
\end{figure}
\begin{figure}[!h]
 \centering
  \includegraphics[width=2.65 in]{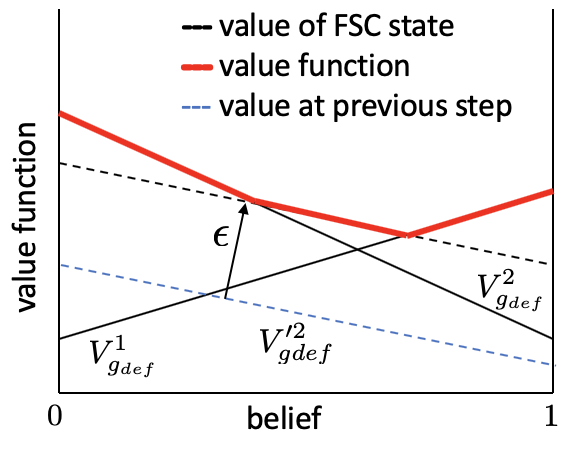} 
  \caption{Robust linear program for state $2$ of $\mathcal{C}_{d}$. Improved vector $V'^2_{g_{d}} + \epsilon$ is tangent to the one-step look-ahead value function. } \label{RobLPOptFiga}
  \end{figure}
%
%A representation for a two-state POSG with three states in $\mathcal{C}_{d}$ is shown in Figure \ref{ValFuncFig}. 
Figure \ref{ValFuncFig} shows $V(b)$ for a two-state POSG with $|G_{d}|=3$.

When working with defender and adversary FSCs of fixed sizes, the value iteration in Algorithm \ref{algo1a} terminates when a (local) equilibrium is reached. 
This means there is no choice of transition probabilities in $\mathcal{C}_{d}$ that will improve the satisfaction probability for some belief state(s). 
This probability can be improved by adding states to $\mathcal{C}_{d}$ (since $|G_{a}|$ is fixed). 
%Notice that the value of a state in $\mathcal{C}_{d}$ is a piece-wise linear function of the belief state. 
The value of a belief $b$ (Equation (\ref{ValForBelief})) is the point-wise maximum of the value at each state of $\mathcal{C}_{d}$, which themselves are linear functions of the belief state. 
Therefore, at equilibrium, 
%the (one-step look-ahead) value function 
$V(b)$ will satisfy: 
\begin{align}
V(b)&=\max_{u_{d}} \min_{u_{a}}\sum_{o \in \mathcal{O}_{d}} \mathbb{P}(o|b)V(b^{u_{d}u_{a}}_{o}) \label{DPBackup}\\
\text{where }\mathbb{P}(o|b) &:= \sum_{s} O_{d}(o|s)b(s),  \label{ObsGivenBelief}\\
%b^{u_{d}u_{a}}_{o}(s')&:=\sum_s \mathbb{T}(s'|s,u_{d},u_{a})\nonumber \\&\quad \quad \quad \quad \times \frac{O_{d}(o|s)b(s)}{\sum_{o \in \mathcal{O}_{d}O_{d}(o|s)b(s)}}\label{LookAheadBelief}
b^{u_{d}u_{a}}_{o}(s'):=\sum_s &\mathbb{T}(s'|s,u_{d},u_{a}) \frac{O_{d}(o|s)b(s)}{\sum_{o \in \mathcal{O}_{d}O_{d}(o|s)b(s)}}\label{LookAheadBelief}
\end{align}
This set of equations is not easy to solve since the belief takes values in $[0,1]$. 
However, Equation (\ref{DPBackup}) results in a point-wise improvement of the value function, until an optimum is reached. 
We will need the following definitions.
\begin{df}[Tangent FSC State]\label{TangFSC}
An FSC state $g_{d}$ is \emph{tangent} to the one-step look-ahead value function $V(b)$ in Equation (\ref{DPBackup}) if $V(b) = V_{g_{d}}(b)$ at state $b$. 
\end{df}
%
%For fixed $G_{d}$, the only way to increase $V_{g_{d}}$ is by adjusting the transition probabilites $\mu_{d}$. 
%
\begin{df}[Improved FSC State]\label{ImprFSC}
A state $g_{d} \in G_{d}$ is \emph{improved} if transition probabilities associated with that state are changed in a way that increases $V_{g_{d}}$. 
\end{df}

For the setting where there are two agents with competing objectives, the problem of determining a policy $\mu_{d}$ that achieves an improvement in $V_{g_{d}}$ under any adversary policy $\mu_{a}$ can be posed as a robust linear program \cite{ben2009robust}, %. 
%For the setting when there are two agents with competing objectives, this problem can be posed as a \emph{robust linear program} (RLP) \cite{ben2009robust} that determines a policy $\mu_{d}$ that achieves an improvement in $V_{g_{d}}$ under any adversary policy $\mu_{a}$. 
%The RLP is 
presented in Equations (\ref{RobustLP})-(\ref{ProbConstr2}).%  Equations (\ref{ProbConstr1})-(\ref{ProbConstr2}) in the robust LP  ensure that $\mu_*(\cdot)$ is a probability distribution. 
%\begin{figure}[h]
% \centering
%  \includegraphics[width=2.85 in]{TangFSC.png} 
%\caption{Equation (\ref{DPBackup}) results in a point-wise improvement of the value function. At some belief states, an improvment is no longer possible. The FSC state $T$ is \emph{tangent} to the one-step look-ahead value function.} \label{TangFSCFig}
%\end{figure}
\begin{figure*}[h]
% ensure that we have normalsize text
\normalsize
% Store the current equation number.
%\setcounter{mytempeqncnt}{\value{equation}}
% Set the equation number to one less than the one
% desired for the first equation here.
% The value here will have to changed if equations
% are added or removed prior to the place these
% equations are referenced in the main text.
%\setcounter{equation}{5}
\begin{align}
\max_{\epsilon, \mu_{d}} &\quad  \epsilon \label{RobustLP}\\
\text{subject to:}&\quad 
V_{g_{d}}(\mathfrak{s}) + \epsilon \leq \sum_{\mathfrak{s}',o,o',g'_{d},g'_{a},u_{d},u_{a}} \Bigg( V_{g'_{d}}(\mathfrak{s}')O_{d}(o|s)\mu_{d}(g'_{d},u_{d}|g_{d},o)O_{a}(o'|s)\nonumber\\& \qquad \qquad \qquad \qquad \qquad \qquad \qquad \times \mu_{a}(g'_{a},u_{a}|g_{a},o')\times \mathbb{T}^{\phi}((s',q')|(s,q),u_{d},u_{a})\Bigg) \nonumber \\
& \text{\hspace{85mm}} \forall s, \forall \mu_{a}(g'_{a},u_{a}|g_{a},o')\\
&\quad  \sum_{g',u_*} \mu_*(g'_*,u_*|g_*,o_*) = 1 \text{\hspace{48mm}}\forall o_*; * \in \{d,a\}\label{ProbConstr1}\\
&\quad \mu_*(g'_*,u_*|g_*,o_*) \geq 0\text{\hspace{55mm}}\forall o_*,g'_*,u_*; * \in \{d,a\}\label{ProbConstr2}
\end{align}
% Restore the current equation number.
%\setcounter{equation}{\value{mytempeqncnt}}
% IEEE uses as a separator
\hrulefill
% The spacer can be tweaked to stop underfull vboxes.
%\vspace*{2pt}
\end{figure*}
%\begin{figure}[!h]
% \centering
%  \includegraphics[width=2.85 in]{RobLPOpt1.png} 
%  \caption{Robust linear program for state $2$ of $\mathcal{C}_{d}$. Improved vector $V'^2_{g_{d}} + \epsilon$ is tangent to the one-step look-ahead value function. } \label{RobLPOptFiga}
%  \end{figure}
%
  \begin{figure}[!h]
  \centering
  \includegraphics[width=2.45 in]{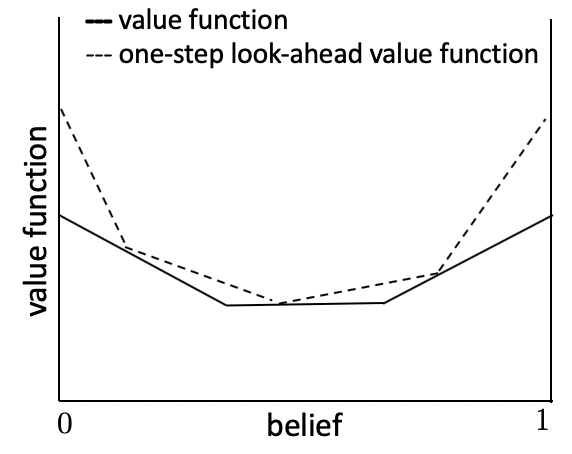} 
  \caption{At a local equilibrium, all states are tangent to the one-step look-ahead value function.} \label{RobLPOptFigb}
\end{figure}

When $\epsilon > 0$, an improvement in the value of the FSC state (by $\epsilon$) can be achieved. 
This is because there exists a convex combination of value vectors of the one-step look-ahead value function that dominates the present value of the FSC state \cite{poupart2004bounded}. 
%$\sum_i b_i V_{g_{d}}(\mathfrak{s}_i)+ \epsilon \geq \sum_i b_i V_{g'_{d}}(\mathfrak{s}_i)$ for some $g'_{d} \in G_{d}$. 
The procedure is carried out for each $g_{d} \in G_{d}$, until no further improvement in the transition probabilities in $\mathcal{C}_{d}$ is possible. 
At this stage, the robust linear program yields $\epsilon = 0$ for every $g_{d} \in G_{d}$. 
The following result generalizes Theorem $2$ in \cite{poupart2004bounded} to a partially observable environment that includes an adversarial agent.
%where it was presented for solving a linear program related to the value of an FSC state for POMDPs. 
%
%
\begin{prop}\label{PropBddFSC}
The policy iteration procedure has reached a local equilibrium if and only if all the states $g_{d} \in G_{d}$ are tangent to $V(b)$.%the one-step look-ahead value function $V(b)$. 
\end{prop}
\begin{IEEEproof}
The robust LP aims to maximize the improvement that can be achieved in the value of each state in $\mathcal{C}_{d}$. 
From the preceding discussion, and from Definition \ref{TangFSC}, a translation of the value vector of a state $g_{d}$ by $\epsilon > 0$ will make it tangent to the one-step look-ahead value function. 
By a similar argument, $\epsilon = 0$ for each $g_{d} \in G_{d}$ indicates that improvement in the value of an FSC state will not be possible if it is already tangent to the one-step look-ahead value function. 
This is shown in Figures \ref{RobLPOptFiga} and \ref{RobLPOptFigb}. 
\end{IEEEproof}

When $\epsilon = 0$ for each $g_{d}$, the satisfaction probability can be improved by adding states to $\mathcal{C}_{d}$ in a `principled way'. 
Let $MaxNewStates$ denote the maximum number of states that can be added to $\mathcal{C}_{d}$, and let $\{b_k\}:=B$ denote the set of belief states satisfying $V(b_k) = V_{g_{d}}(b)$ for each $g_{d}$ from Equation (\ref{RobustLP}).
\begin{algorithm}[!h]
\renewcommand{\algorithmiccomment}[1]{// #1}
\caption{Adding states to FSC $\mathcal{C}_{d}$}\label{algo1b}
\begin{algorithmic}[1]
\REQUIRE Set of belief states $\{b_k\}:=B$ ; $MaxNewStates$
\ENSURE Set of improved states in FSC $\mathcal{C}_{d}$
\STATE $NewStateNumber \leftarrow 0$
\WHILE {$B \neq \emptyset$}
\STATE Choose $b \in B$
\STATE $B:=B \setminus b$
\STATE $Ahead : = \emptyset$
\FOR {$(u_{d},u_{a}, o_{d}) \in U_{d} \times U_{a} \times \mathcal{O}_{d}$}
\IF {$\mathbb{P}(o|b) > 0$ in Equation (\ref{ObsGivenBelief})}
\STATE Determine $b^{u_{d}u_{a}}_{o}(s')$ from Equation (\ref{LookAheadBelief})
\STATE $Ahead = Ahead \cup \{b^{u_{d}u_{a}}_{o}\}$
\ENDIF
\ENDFOR
\FOR {$b_a \in Ahead$}
\STATE Determine $V(b_a)$ from Equations (\ref{DPBackup}, \ref{ObsGivenBelief}, \ref{LookAheadBelief})
\STATE Note maximizers $u^*_{d}$, $g^*_{d}$ (Eqns. (\ref{DPBackup}),  (\ref{ValForBelief}))
\IF {$V(b_a) > V(b)$ \textbf{AND} $NewStateNumber < MaxNewStates$}
\STATE Add state, $g_{new}$ to $\mathcal{C}_{d}$ with $\mu_{d}(g^*_{d}, u^*_{d}|g_{new}, o) = 1$ $\forall o \in \mathcal{O}_{d}$
\STATE $NewStateNumber \leftarrow NewStateNumber + 1$
\ENDIF
\ENDFOR
\ENDWHILE
\end{algorithmic}
\end{algorithm}

Algorithm \ref{algo1b} presents a procedure to add states to the defender FSC to improve the satisfaction probability. 
%Moreover, the procedure terminates in finite time, which is shown in Proposition \ref{PropFinTime}. 
Lines $6-11$ determine the one-step look-ahead beliefs. 
A new state is added to $\mathcal{C}_{d}$ if the defender `believes' that the probability of satisfying the LTL formula from these states is higher than that from the current belief state (Lines $13-18$). 
%This is shown in Lines $13-18$. 
Lines $13$ and $14$ respectively form the policy evaluation and policy improvement steps of the policy iteration. 
Edges from new states in the FSC are directed towards the action and FSC state that maximize the value function $V(b)$ in a deterministic manner (Line $16$). 
%Note that this is only a heuristic. 
%Transitions to and from the added states and the associated probabilities will be adjusted as other FSC states are improved.
Values of probabilities and transitions to and from the added state will be adjusted as other FSC states are improved. 
\begin{prop}\label{PropFinTime}
Algorithm \ref{algo1b} terminates in finite time.
\end{prop}
\begin{IEEEproof}
This follows from the facts the sets $B$, $\mathcal{O}_{d}$, $U_{d}$, and $U_{a}$ have finite cardinality, and the one-step look-ahead values in Equation (\ref{DPBackup}) are upper bounded. 
\end{IEEEproof}

The procedure yields a new $\mathcal{C}_{d}$, from which candidate FSC structures can be found using Algorithm \ref{algo1}.  
The defender policy to maximize the probability of satisfying the LTL formula under any $\mathcal{C}_{a}$ of fixed size can be determined by Algorithm \ref{algo1a} or the robust linear program in Eqn. (\ref{RobustLP}).
\section{Examples}\label{Example}

This section presents an example and experiments that illustrate our approach. 
\begin{figure}[!h]
 \centering
  \includegraphics[width=2.45 in]{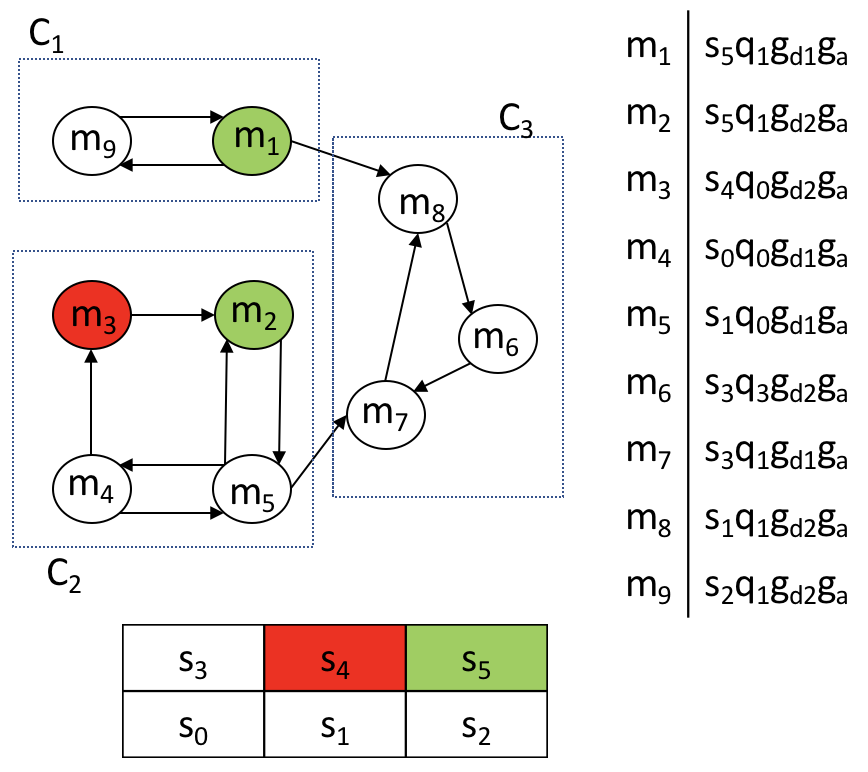} 
\caption{Clockwise, from \emph{top-left}: Global Markov chain (GMC) for initial defender and adversary FSC structures- green states ($m_1 \& m_2$) must be visited infinitely often, and state in red ($m_3$) must be visited finitely often in steady-state; GMC state $m_i \in S \times Q \times G_{d} \times G_{a}$; State-space for $M=3, N=2$ showing unsafe ($s_4$) and target ($s_5$) states.}\label{GMCInit}
\end{figure}
\begin{eg}\label{IllustrEg}

For this example, the state space is an $M \times N$ grid, $S:=\{s_i:i = x+My, x \in \{0,\dots,M-1\}, y \in \{0, \dots, N-1\}\}$. 
%This will define an $M \times N$ grid. 
%
The defender's actions are $U_{d} = \{R, L, U, D\}$ denoting right, left, up, and down, and the actions of the adversary are $U_{a} = \{A, NA\}$, denoting attack, and not attack respectively. 
The observations of both agents are $\mathcal{O}_{d} = \mathcal{O}_{a} = \{correct, wrong\}$, such that:
%\begin{align*}
$O_{d}(correct|s_i) = 0.8 = 1-O_{d}(wrong|s_i)$, and 
$O_{a}(correct|s_i) = 0.6 = 1-O_{a}(wrong|s_i)$.
%\end{align*} 
%
$\mathcal{O}_{d}$ and $ \mathcal{O}_{a}$ are probabilities that the agents sense that their observation of the state is indeed the correct state or not. 
That is, $\mathbb{P}(o_i = s_i)$ or $\mathbb{P}(o_i \neq s_i)$. 
  \begin{figure*}
  \centering
  \begin{subfigure}[t]{0.48\textwidth}
  \centering
  \includegraphics[width = 0.87 \linewidth]{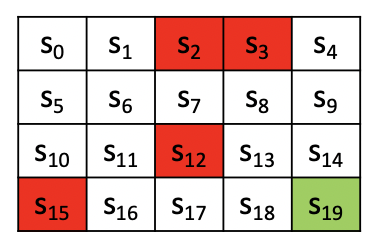} 
  %\caption{Grid-world with $20$ states. States in red indicate the presence of an obstacle, and the state in green is the target state.} 
  \caption{}\label{ExptFSCGrid}
  \end{subfigure}
  \begin{subfigure}[t]{0.48\textwidth}
  \centering
  \includegraphics[width = 0.99 \linewidth]{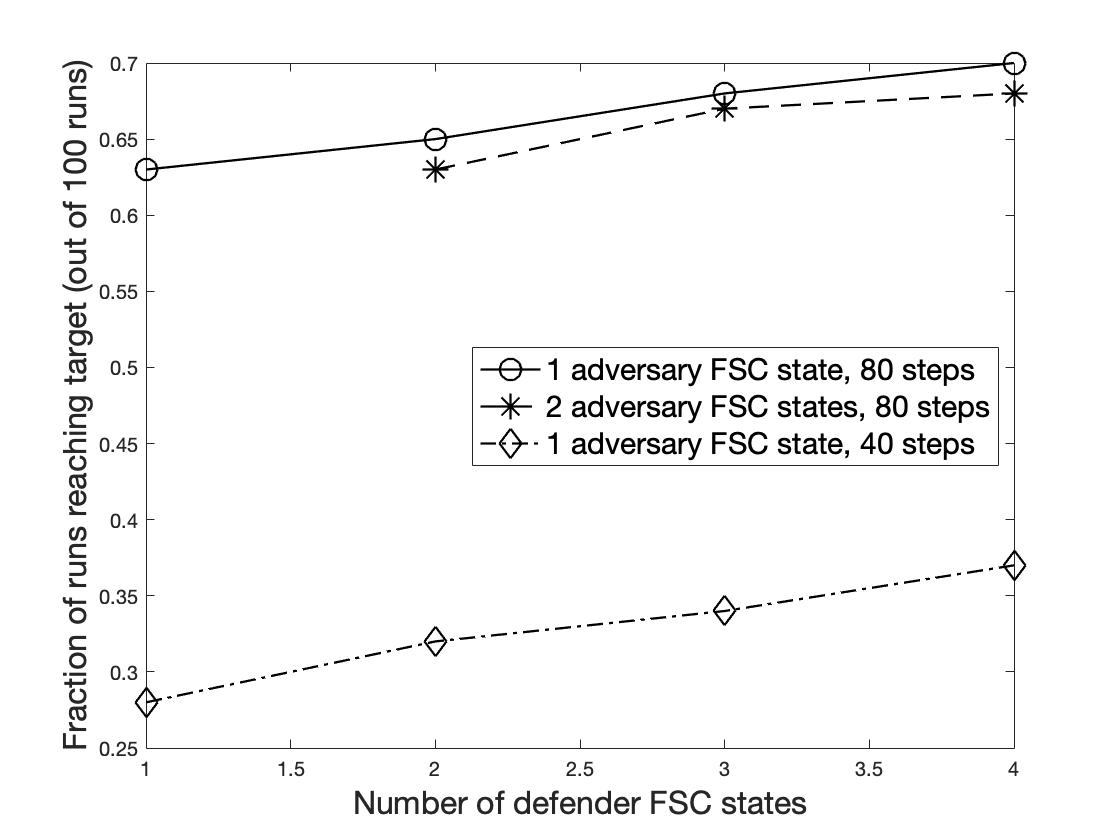} 
  %\caption{Fraction of runs (out of $100$) when agent reaches target within $80$ steps (blue, red curves), and $40$ steps (green curve).} 
  \caption{}\label{ExptFSCGraph}\label{ExptFSCGraph}
  \end{subfigure}
 \caption{The agent aims to satisfy the LTL formula $\phi = \mathbf{G}\mathbf{F} \mathtt{tar} \wedge \mathbf{G} \neg \mathtt{obs}$ in the presence of an adversary, in a partially observable environment. The environment is the grid-world in Figure \ref{ExptFSCGrid}. The states in red indicate the presence of an obstacle, the state in green is the target state, and the agent starts in state $s_0$. The agents' actions are determined by their observations of the state. Assume that the defender FSC has at least as many states as the adversary FSC, i.e. $|G_{d}| \geq |G_{a}|$. Figure \ref{ExptFSCGraph} shows the fraction of runs (out of $100$) when the agent reaches the target within $80$ steps and $40$ steps. This number is higher when $|G_{d}|$ is larger, for a fixed value of $|G_{a}|$ ($-\mathtt{o}-, -*-$, and $-\Diamond-$ curves). For the same $|G_{d}|$, the fraction of successful runs is higher when $|G_{a}|$ is lower ($-\mathtt{o}-$ and $-*-$ curves). The fraction of successful runs is also higher when the agent is allowed more steps to reach the target ($-\mathtt{o}-$ and $-\Diamond-$ curves).}\label{FigExpt}
\end{figure*}

Let $\mathcal{AP} = \{\mathtt{obs}, \mathtt{tar}\}$, denoting obstacle and target respectively. 
Then, if $\phi = \mathbf{G}\mathbf{F} \mathtt{tar} \wedge \mathbf{G} \neg \mathtt{obs}$, the corresponding DRA will have two states $q_0, q_1$, with $F = (\{\emptyset\}, \{q_1\})$. 
Transition probabilities for $(u_{d}, u_{a}) = (R, NA)$ and $(R, A)$ are defined below. 
Let $N_{s_i}$ denote the neighbors of $s_i$.
%
%\begin{align*}
%\mathbb{T}(s_j|s_i, R, NA)=
%\begin{cases}
%0.8 & \text{$j=i+1$, $(i+1)\not\equiv 0 \bmod M$}\\ 
%\frac{0.2}{|N_{s_i}|} & \text{($s_j \in  \{s_i\} \cup N_{s_i}\setminus \{s_{i+1}\}$), $(i+1) \not\equiv 0 \bmod M$}\\
%1 & \text{$j =i$ and $(i+1) \equiv 0 \bmod M$}\\
%\end{cases}
%\end{align*}
%\begin{align*}
%\mathbb{T}(s_j|s_i, R, A)=
%\begin{cases}
%0.6 & \text{$j=i+1$, $(i+1)\not\equiv 0 \bmod M$}\\ 
%\frac{0.4}{|N_{s_i}|} &\text{($s_j \in  \{s_i\} \cup N_{s_i}\setminus \{s_{i+1}\}$), $(i+1) \not\equiv 0 \bmod M$}\\
%1 & \text{$j =i$ and $i+1  \equiv 0 \bmod M$}
%\end{cases}
%\end{align*}
%
\[
\mathbb{T}(s_j|s_i, R, NA)=
\begin{cases}
0.8 & \text{$j=i+1$, $(i+1)\not\equiv 0 \bmod M$}\\ 
\frac{0.2}{|N_{s_i}|} & \parbox{0.20 \textwidth}{($s_j \in  \{s_i\} \cup N_{s_i}\setminus \{s_{i+1}\}$), $(i+1) \not\equiv 0 \bmod M$}\\
1 & \text{$j =i$ and $(i+1) \equiv 0 \bmod M$}\\
\end{cases}\]
\[
\mathbb{T}(s_j|s_i, R, A)=
\begin{cases}
0.6 & \text{$j=i+1$, $(i+1)\not\equiv 0 \bmod M$}\\ 
\frac{0.4}{|N_{s_i}|} &\parbox{0.20 \textwidth}{($s_j \in  \{s_i\} \cup N_{s_i}\setminus \{s_{i+1}\}$), $(i+1) \not\equiv 0 \bmod M$}\\
1 & \text{$j =i$ and $i+1  \equiv 0 \bmod M$}
\end{cases}
\]

Notice that in the above equations, the probability of the agent moving to the `correct' next state for a particular defender action is larger for the adversary action $NA$ than for the adversary action $A$. 
Further, in this case, if the defender is in a square along the right edge of the grid, then the action $R$ does not result in a change of state. 
The probabilities for other action pairs can be defined similarly.

For this example, let $M = 3$ and $N = 2$. 
Then, $|S| = 6$. 
Let $s_4$ be an unsafe state, and $s_5$ be the goal state. 
This is indicated in Figure \ref{GMCInit}. 
Let $|G_{d}| = 2$, $|G_{a}|=1$ for the FSCs. 
Assume that for some initial structures $\mathcal{I}_{d}^0, \mathcal{I}_{a}^0$ the GMC is given by Figure \ref{GMCInit}. 
The figure also indicates the states in terms of its individual components. 
Assume that the LTL formula $\phi$ is such that the states in green denote those that have to be visited infinitely often in steady state, while those in red must be avoided. 
Therefore $(L^{\phi}, K^{\phi}) = \{(\{\emptyset\}, \{m_1\}), (\{m_3\},\{m_2\})\}$. 
The boxes $C_1, C_2, C_3$ indicate the communicating classes of the graph. 

From Algorithm \ref{algo1}, for $C_1$, $Bad = \{m_8\}, Good = \{m_1\}$. 
For $m_1 \rightarrow m_8$, Eqn. (\ref{BigEqn1}) is true for all $u_{a}$ and $u_{d} = \{D, L\}$. 
Thus, $\mathcal{I}_{d}(g',u_{d}|g,o) \leftarrow 0$ for $o = \{correct, wrong\}$. 
For $m_9 \rightarrow m_1$, since Eqn. (\ref{BigEqn2}) does not hold for $R, D \in U_{d}$, $\mathcal{I}_{a}(\cdot)$ is unchanged. 
Then, $m_1$ is recurrent in $\mathcal{G}_{new}$. 
For $C_2$, $Bad = \{m_3, m_7\}, Good = \{m_2\}$. 
Like for $C_1$, $\mathcal{I}_{a}(\cdot)$ remains unchanged, since (\ref{BigEqn2}) does not hold for $D \in U_{d}$. 
For $m_5 \rightarrow m_7$, $\mathcal{I}_{d}(g',u_{d}|g,o) \leftarrow 0 \forall u_{d} \in U_{d}\setminus D$. 
A similar conclusion is drawn for $m_4 \rightarrow m_3$. 
Then, $m_2$ will be recurrent in $\mathcal{G}_{new}$. 
For $C_3$, since $Bad = Good = \emptyset$, no structure is added to $\mathcal{I}$. 
Notice that these FSCs satisfy Proposition \ref{Proposn}. 

This example also shows the limitations of Algorithm \ref{algo1}. 
From the $M \times N$ grid, there is a policy that takes the defender from any $s \in S \setminus \{s_4\}$ to $s_5$ with probability $1$. However, for FSCs of small size, the initial state of the defender might result in the Algorithm reporting that no solution was found, even if there exists a feasible solution.
\end{eg}
\begin{eg}
Consider the model of Example \ref{IllustrEg}, with $M=5$, $N=4$. 
A representation of the environment is shown in Figure \ref{ExptFSCGrid}. 
Like in Example \ref{IllustrEg}, the LTL formula to be satisfied is $\phi = \mathbf{G}\mathbf{F} \mathtt{tar} \wedge \mathbf{G} \neg \mathtt{obs}$. %, where $\mathtt{tar}$ and $\mathtt{obs}$ respectively denote the \emph{target} and \emph{obstacles}. 
The observation function of the defender is modified so that for a state $s$ where $\mathcal{L}(s) = \mathtt{obs}$ or $\mathcal{L}(s) = \mathtt{tar}$, $O_{d}(correct|s) = 1$. 
That is, the defender recognizes an obstacle or the target correctly with probability one. 
Our experiments compute the probability of reaching the target under limited sensing capabilities of the agents with FSCs having different number of states. 
In each case, we assume that $|G_{d}| \geq |G_{a}|$. 
The number of states in the GMC varies from $60$ to $480$. 
%The environment and experimental results are shown in Figure \ref{FigExpt}. 

%The graph only 
\begin{table}[!h]
\begin{center}
\begin{tabular}{|c|c|c|} \hline
$\mathbf{|G_d|}$ & $\mathbf{|G_a|=1}$: $\mathbf{V(\mathfrak{s}_0)}$ \textbf{[Std. Dev.]} & $\mathbf{|G_a|=2}$: $\mathbf{V(\mathfrak{s}_0)}$ \textbf{[Std. Dev.]} \\ \hline 
$1$      & $0.53$ [$0.04$]                          & $0.45$ [$0.05$]         \\ \hline
$2$      & $0.55$ [$0.05$]                          & $0.45$ [$0.06$]         \\ \hline
$3$      & $0.56$ [$0.09$]                           & $0.47$ [$0.08$]         \\ \hline 
$4$      & $0.57$ [$0.10$]                          & $0.48$ [$0.12$]        \\ \hline 
\end{tabular}
\caption{Satisfaction probability and standard deviation (over 100 trials) of reaching the target state $s_{19}$ from $s_0$ for LTL formula $\phi = \mathbf{G}\mathbf{F} \mathtt{tar} \wedge \mathbf{G} \neg \mathtt{obs}$ starting from $s_0$ for varying number of defender FSC states $|G_d|$, when number of states in adversary FSC, $|G_a| = 1$ and $|G_a|=2$.}\label{TableStdDev}
\end{center}
\end{table}
Table \ref{TableStdDev} shows the average satisfaction probability and standard deviation of reaching the target state $s_{19}$ starting from $s_0$ (expressed in terms of the value of the state from Equation (\ref{SatProb})) when the adversary FSC has one and two states. 
Higher values of the standard deviation could be due to the fact that in some cases, Algorithm \ref{algo1a} may terminate before $V(\mathfrak{s}_0)$ is updated enough number of times. 

\begin{table}[!h]
\begin{center}
\begin{tabular}{|c|c|c|c|} \hline
$\mathbf{|G_d|}$ & \textbf{Benign Baseline} & \textbf{Adv. Baseline} & \textbf{Adv.-Aware (ours)}\\ \hline 
$1$      & $0.69$                           & $0.35$     & $\mathbf{0.53}$    \\ \hline
$2$      & $0.70$                          & $0.35$    & $\mathbf{0.55}$   \\ \hline
$3$      & $0.73$                      & $0.38$    & $\mathbf{0.56}$     \\ \hline 
$4$      & $0.75$                     & $0.39$    & $\mathbf{0.57}$    \\ \hline 
\end{tabular}
\caption{Comparison of probabilities of satisfying LTL formula $\phi = \mathbf{G}\mathbf{F} \mathtt{tar} \wedge \mathbf{G} \neg \mathtt{obs}$ starting from $s_0$ in the presence and absence of an adversary. The first column lists the number of defender FSC states. Subsequent columns enumerate satisfaction probabilities in the following scenarios: i) absence of adversary (Benign Baseline \cite{sharan2014formal}); ii) using a defender policy that was synthesized without an adversary, but realized in the presence of an adversary (Adversarial Baseline); iii) using a defender policy designed assuming the presence of an adversary (Adversary-Aware Design- \textbf{our approach}). Although the benign baseline gives the highest satisfaction probability, the same baseline when used in the presence of an adversary results in a much lower satisfaction probability. In comparison, our \emph{Adversary-Aware Design} approach results in a higher satisfaction probability than the `Adversarial Baseline' where we use a defender policy designed to account for adversarial behavior. We assume that the adversary FSC has one state, so that the GMC with and without the adversary FSC will have the same number of states.}\label{TableCompBaseline}
\end{center}
\end{table}
Table \ref{TableCompBaseline} compares the probabilities of satisfying the LTL objective $\phi = \mathbf{G}\mathbf{F} \mathtt{tar} \wedge \mathbf{G} \neg \mathtt{obs}$ starting from $s_0$ in the presence and absence of an adversary. 
We compare our approach with a baseline, which is a defender policy synthesized in the absence of an adversary. 
The GMC for the case without an adversary is constructed using the approach of \cite{sharan2014formal}. 
This baseline defender policy in then realized in the presence of an adversary, and we compare it with our method of synthesizing a defender policy assuming the presence of an adversary. 
Although the highest satisfaction probability is got while using the baseline policy, when this baseline is used in the presence of an adversary, we obtain a much lower satisfaction probability. 
In comparison, our \emph{adversary-aware defender policy} results in a higher satisfaction probability than when using the baseline in the presence of the adversary.
We note that for this comparison, the adversary FSC has one state, i.e., $|G_a| = 1$, so that the GMCs with and without the adversary FSC have the same number of states.

Figure \ref{ExptFSCGraph} shows the fraction of sample paths when the agent reaches the target for the first time. 
After this, since the agent is in a recurrent set, it will continue to visit states in this set with probability one.
The following observations can be drawn from Figure \ref{ExptFSCGraph}. 
First, for a fixed $|G_{a}|$, the fraction of runs when the agent successfully reaches the target increases as $|G_{d}|$ increases. 
Second, for a fixed $|G_{d}|$, the probability of satisfying $\phi$ is higher for a smaller $|G_{a}|$. 
Third, the fraction of successful runs improves with allowing the agent more steps to reach the target. 
One reason for the first two observations is that the number of states in an FSC models the `memory' available to the agent. 
The defender can play better when it has more FSC states, or when the adversary has fewer FSC states. 
While our results agree with intuition, a caveat is that these numbers depend on the agents' observations, $\mathcal{O}_{d}$ and $\mathcal{O}_{a}$. 
%In this example, 
Here, the observations of the agents is an indication of whether the state is the actual state the defender is in. % or not. 
This could be a reason for fewer successful runs when the agent is allowed a maximum of $40$ steps versus the case when it is allowed $80$ steps.
\end{eg}
\section{Related Work}\label{RelWork}

A large body of work studies classes of problems that are relevant to this paper. 
These can be divided into three broad categories: \emph{i)} synthesis of strategies for systems represented as an MDP that has to additionally satisfy a TL formula; \emph{ii)} synthesis of strategies for POMDPs; \emph{iii)} synthesis of defender and adversary strategies for an MDP under a TL constraint. 
While there has been recent work on the synthesis of controllers for POMDPs under TL specifications, these have largely been restricted to the single-agent case, and do not address the case when there might be an adversary with a competing objective. 
 
Approaches that address the satisfaction of TL constraints for problems in motion-planning include hierarchical control \cite{fainekos2009temporal}, ensuring probabilistic satisfaction guarantees \cite{lahijanian2012temporal}, and sensing-based strategies \cite{kress2007s}. 
Controller synthesis for deterministic linear systems to ensure that the closed-loop system will satisfy an LTL formula is studied in \cite{kloetzer2008fully}. 
The authors of \cite{wolff2012robust} propose methods to synthesize a robust control policy that satisfies an LTL formula for a system represented as an MDP whose transitions are not exactly known, but are assumed to lie in a set. 
For MDPs under an LTL specification, a partial ordering on the states is leveraged to solve controller synthesis as a receding horizon problem in \cite{wongpiromsarn2012receding}. 
The synthesis of an optimal control policy that maximizes the probability of an MDP satisfying an LTL formula that additionally minimizes the cost between satisfying instances is studied in \cite{ding2014optimal}. 
This is computed by determining \emph{maximal end components} in an MDP. 
However, this approach will not work in the partially observable setting, where policies will depend on an observation of the state \cite{sadigh2014learning}. 
The synthesis of joint control and sensing strategies for discrete systems with incomplete information and sensing is presented in \cite{fu2016synthesis}. 
The setting of \cite{ding2014optimal} in the presence of an adversary with competing objectives was presented in \cite{niu2018optimal}.

A policy in a POSG (or POMDP) %depends on the `history' of the system. 
%That is, a policy 
at time $t$ depends on actions and observations at all previous times. 
A memoryless policy, on the other hand, only depends on the current state. 
For fully observable stochastic games, it is possible to always find memoryless policies that are optimal. 
However, a policy with memory could perform much better than a memoryless policy for POSGs.
One way of determining policies for a POSG is to keep track of the entire execution, observation, and action histories, which can be abstracted into determining a sufficient statistic for the POSG execution. 
%An example of a sufficient statistic that has widely been studied in the POMDP literature is the \emph{belief state}, an abstraction that reflects the probability that the agent is in a state based on receiving observations from the environment. 
One example is the \emph{belief state}, which reflects the probability that the agent is in some state, based on receiving observations from the environment. 
Updating the \emph{belief state} at every time step only requires knowledge of the previous belief state and the most recent action and observation. 
Thus, the belief states form the states of an MDP \cite{smallwood1973optimal}, which is more amenable to analysis \cite{bertsekas2015dynamic} than a POMDP.   
However, the belief state is uncountable, and will hinder development of exact algorithms to determine optimal strategies.% since these will require nontrivial and potentially infinite memory.

Synthesis of memoryless strategies for POMDPs in order to satisfy a specification was shown to be NP-hard and in PSPACE in \cite{vlassis2012computational}. 
In \cite{wongpiromsarn2012control}, a discretization of the belief space was carried out \emph{apriori}, resulting in a fully observable MDP. 
However, this approach is not practical if the state space is large \cite{yu2008near}. 
The complexity of determining a strategy for maximizing the probability of satisfaction of %a broad class of objectives called 
parity objectives was shown to be undecidable in \cite{chatterjee2013survey}. 
However, determining finite-memory strategies for the qualitative problem of parity objective satisfaction was shown to be EXPTIME-complete in \cite{chatterjee2014complexity}.

Dynamic programming for POSGs for the finite horizon setting was studied in \cite{hansen2004dynamic}. 
%, resulting in an algorithm that generalizes both DP for POMDPs and iterated elimination of dominated strategies for normal form games.
%This work, however, considered the finite horizon case, and all agents had to maximize their own expected rewards. 
When agents cooperate to earn rewards, the framework is called a decentralized-POMDP (Dec-POMDP). 
The infinite horizon case for Dec-POMDPs was presented in \cite{bernstein2005bounded}, where the authors proposed a bounded policy iteration algorithm for policies represented as joint FSCs. 
A complete and optimal algorithm for deterministic FSC policies for DecPOMDPs was presented in \cite{szer2005optimal}. 
Optimization techniques for `fixed-size controllers' to solve Dec-POMDPs were investigated in \cite{amato2010optimizing}. 
A survey of recent research in Dec- POMDPs is presented in \cite{oliehoek2016concise}.

The satisfaction of an LTL formula in partially observable adversarial environments was presented for the first time in \cite{bhaskar2019finite}. 
The authors used FSCs to represent the policies of the agents, and proposed an algorithm that yielded defender FSCs of fixed size satisfying the LTL formula under any adversary FSC of fixed size. 
The authors also broadened the scope of this problem to continuous state environments, settings when the adversary could potentially tamper with clocks that keep track of time in the environment, and where an additional privacy constraint had to be satisfied in \cite{ramasubramanian2019linear, niu2020controlmetric, ramasubramanian2020privacy}.
\section{Conclusion}\label{Conclusion}

This paper demonstrated the use of FSCs in order to satisfy an LTL formula in a partially observable environment, in the presence of an adversary. 
The FSCs represented agent policies, and these were composed with a POSG representing the environment to yield a fully observable MC. 
We showed that the probability of satisfaction of the LTL formula was equal to the probability of reaching recurrent classes of this MC. 
We subsequently presented a procedure to determine defender and adversary controllers of fixed sizes that result in nonzero satisfaction probability of the LTL formula, and proved its soundness. 
Maximizing the satisfaction probability was related to reaching a Stackelberg equilibrium of a stochastic game involving the agents through a value-iteration based procedure. 
Finally, we showed a means to add states to the defender FSC in a principled way in order to improve the satisfaction probability for adversary FSCs of fixed sizes. 

In Section \ref{FSCVary}, when adding states to $\mathcal{C}_{d}$, we assumed that the size of $\mathcal{C}_{a}$ was fixed. 
Future work will seek to relax this assumption, and study cases when the defender has limited knowledge of the abilities of an adversary. 
%We will also study methods to quantitatively compare the optimality of FSC-based policies with policies determined using other heuristics to approximately solve partially observable environments.
%
%A major challenge in solving problems of the kind studied in this paper is that of \emph{state explosion}. 
%A direction of future work is to study state aggregation techniques \cite{ren2002state, li2006towards, clarke2011model} for the product-POSG and the GMC, which will enable solutions of more complex problems. 
To address the challenge of \emph{state explosion}, we will investigate state aggregation techniques \cite{ren2002state, li2006towards, clarke2011model} for the product-POSG and the GMC. This will enable solutions of more complex problems. 
%The formal methods research community has developed several methods to circumvent this problem, including ordered binary decision diagrams, counter-example guided abstraction refinement, and bounded model checking. 
%A survey of some of these techniques is presented in \cite{clarke2011model}. 
%On the other hand, there has been parallel effort in developing efficient state aggregation procedures for MDPs \cite{ren2002state, li2006towards}, where groups of states are treated as one by ignoring irrelevant state information. 
%The problem is solved on the smaller aggregated state space, and guarantees are provided when the solution is translated to the original, larger state space. 
%Since the framework in this paper involves both these areas, it will be of interest to study state aggregation techniques for the product-POSG and the GMC, and will potentially allow us to solve more complex problems. 
%
%Another area of interest is to use reinforcement learning (RL) when the environment is not known (that is, the transition probabilities of the MDP are unknown). 
%For the single-agent case, determining a policy so that an LTL formula is satisfied in an unknown environment using RL has been studied in \cite{chen2013temporal, li2018policy}
%A preliminary approach for unknown adversarial environments with the agents exactly observing the state was presented in \cite{muniraj2018enforcing}. 
%%We note that the aforementioned paper considered a temporal logic different from LTL. 
%A challenge will be to extend these techniques to partially observable unknown environments with an adversary.
%
\bibliographystyle{IEEEtran}
\bibliography{PartObsCont.bib}

\end{document}